\documentclass{article}

\PassOptionsToPackage{numbers}{natbib}
\usepackage{arxiv}

\renewcommand{\undertitle}{}
\renewcommand{\headeright}{}

\usepackage[utf8]{inputenc} 
\usepackage[T1]{fontenc}    
\usepackage{hyperref}       
\usepackage{url}            
\usepackage{booktabs}       
\usepackage{amsfonts}       
\usepackage{nicefrac}       
\usepackage{microtype}      
\usepackage{graphicx}
\usepackage{natbib}
\usepackage{doi}

\usepackage[linesnumbered, ruled]{algorithm2e}

\SetCommentSty{mycommfont}
\SetKwInOut{Input}{Input}
\SetKwInOut{Output}{Output}
\usepackage{algpseudocode}

\usepackage{multirow}
\usepackage{makecell}
\usepackage{subcaption}
\usepackage{float}
\usepackage{placeins}
\usepackage{dblfloatfix}

\usepackage{amsthm}%
\usepackage{amsmath, nccmath}%
\usepackage{makecell}

\newcommand{\SLC}{\textsf{SLC}}

\begin{document}

\title{Neighborhood Watch: Privacy Risks in Seeded Local Combination Synthetic Data}

\author{
Hadrien Lautraite\\
{\normalfont Université du Québec à Montréal}
\And
Tristan Allard\\
{\normalfont Université de Rennes}
\And
Anne-Sophie Charest\\
{\normalfont Université Laval}
\AND
Jean-François Rajotte\\
{\normalfont Université du Québec à Montréal}
\And
Sébastien Gambs\\
{\normalfont Université du Québec à Montréal}
}

\date{}

\maketitle

\begingroup
\renewcommand{\thefootnote}{}
\footnotetext{Corresponding author: Hadrien Lautraite
(\texttt{lautraite.hadrien@courrier.uqam.ca}).}
\endgroup

\begin{abstract}
Synthetic data is seen as a promising solution for sharing data in sensitive contexts.
However, recent work on privacy attacks have shown that there are still significant residual risks, especially for synthetic data generations methods that are not based on formal approaches such as differential privacy. 
In this paper, we investigate the privacy risks associated with local combination approaches for generating synthetic data in which synthetic profiles are built by combining real neighbouring profiles. 
More precisely, we focus on three methods from this family, namely SMOTE, Simulant and Avatar, which have been recently used as a way to share ``anonymised data'' in the healthcare domain. 
In particular, we conduct an extensive privacy analysis through a diverse set of attacks: membership inference, linkage and reconstruction attacks. 
Our results demonstrate substantial privacy leakage for all three methods, raising serious doubts about whether their outputs should be regarded as anonymous in practice.
\end{abstract}



\maketitle

\section{Introduction}

Being able to share sensitive data in domains such as healthcare is fundamental to be able to train machine learning models or conduct large-scale data analysis~\cite{9134853}.
However, directly sharing data such as patient hospital records is generally impossible due to the associated privacy risks.
Several data protection authorities across the world have considered the use of synthetic data as a possible way to share anonymized data~\cite{singapore-guide,ico,ipc,pipc,aepd}.
However, the recent development of privacy attacks tailored to synthetic data has highlighted the residual privacy risks arising from such methods, thus questioning their real privacy guarantees~\cite{stadler2022synthetic}.

In this paper, we investigate the privacy guarantees of synthetic data generation for tabular data. 
While a wide range of methods exists for this purpose, including marginal-based approaches~\cite{mckenna2021winning} and deep learning approaches such as GAN-based~\cite{xu2019modeling}, we consider simpler methods which often perform well for smaller datasets and are easier and quicker to implement, which makes them particularly appealing to practitioners.  
More precisely, we focus on the privacy evaluation of a family of generative modeling techniques that work by first selecting a seed profile and then generating a synthetic profile through a (noisy) combination of real profiles that are neighbours to the seed.
We refer to these methods as \emph{Seeded Local Combination} (\SLC) methods.

A first example of \SLC~method is SMOTE~\cite{chawla2002smote}, commonly used in the healthcare domain as a data augmentation method but also as a synthetic data generation approach~\cite{kaabachi2025scoping}. 
Avatar~\cite{guillaudeux2023patient} is the second considered \SLC~method, which forms the basis of a commercial solution 
used by several hospitals, universities, companies as well as governmental agencies~\cite{octopize2025}. 
In particular, Avatar was recently used to release synthetic placebo profiles from a randomized clinical trial for patient suffering from multiple sclerosis~\cite{demuth2025privacy} as well as for releasing synthetic data in a genetic study of patient suffering from multiple sclerosis~\cite{paris2025genetic}.
These two studies report using Avatar with the privacy parameter $k=2$, which in this setting makes Avatar very similar to SMOTE as both methods produce synthetic profiles by interpolating between two real profiles.
Finally, we also consider as a third method studied, the Simulant \SLC~method~\cite{beigi2023simulants} from 
a company providing software for clinical studies~\cite{medidata2026}.

\emph{While the privacy risks associated with synthetic data have already been studied for well-known methods~\cite{hayes2017logan,wu2025winning,golob2025privacy}, attacks on \SLC~methods have not received as much attention}. 
Our study shows that all studied \SLC~methods are highly vulnerable to membership inference attacks (MIAs) as well as linkage attacks tailored to this setting.
Moreover, we extend a previously proposed reconstruction attack and show that it can for instance reconstruct a significant part of one of the datasets used in our experiments (the WBCD dataset).
Our work thus raises doubts about whether data produced by these \SLC~techniques can be considered anonymous.  
Our main contributions can be summarized as follows.
\begin{enumerate}
    \item We have designed an enhanced ensemble black-box MIA that achieves a TPR@10\%FPR significantly better than random guessing, and that also consistently outperforms single baseline attacks. 
    \item We have developed a novel linkage attack between real and synthetic profiles adapted to \SLC. 
    While previous works rely on the distance between real and synthetic profiles, we model the synthetic data distribution generated from each real profile and use the likelihood that a given synthetic profile arose from each distribution to infer its source profile.
    \item We extend a reconstruction attack previously developed against SMOTE~\cite{ganev2025smotemirrorsexposingprivacy}, called \textsf{ReconSMOTE}, targeting synthetic data used to oversample from a minority class by searching for intersection of lines supporting three synthetic points. 
    We propose an extension in the non-oversampling scenario, which is effective against synthetic data generated by Avatar with privacy parameter $k=2$, a setting used in practice for the release of medical data. 
\end{enumerate}
\emph{Outline.}
First, in Section~\ref{sec: Seeded local combination synthetic data generation} we describe the necessary background on the \SLC~approaches studied within this work.
Then, in Section~\ref{sec:related} we review the related work on privacy attacks and metrics tailored to synthetic data and \SLC. 
Afterwards, we describe the novel attacks that we have developed in Section~\ref{sec_attacks} before reporting on their empirical evaluation on three datasets in Section~\ref{sec:exp_evaluation}. 
Finally, we discuss differential privacy in relation to the \SLC~approaches in Section~\ref{sec:discussion} and offer concluding remarks in Section~\ref{sec:conclusion}.

\section{Seeded local combination synthetic data generation}
\label{sec: Seeded local combination synthetic data generation}
We consider a dataset $D$ composed of $n$ profiles each described by $d$ attributes. 
Anonymization techniques usually aim at modifying the individual profiles to prevent privacy risks such as re-identification attacks.
In contrast, most synthetic data generation approaches learn a generative model $f$ capturing the data distribution $P(D)$, from which a synthetic dataset $D_{synth}$ is then sampled.

\emph{Seeded local combination.} We coin the family of methods that we consider as Seeded Local Combination (\SLC) approaches.
These methods can be abstracted as a three steps process described in Algorithm~\ref{alg:sli_family}. 
First, a profile $x^{seed}$ is selected from dataset $D$ to act as a seed for the synthetic data generation. 
Then, $k$ data points are selected in the neighborhood of $x^{seed}$ to form the set $N$.
Finally, a synthetic data profile $s$ is obtained from a noisy combination of the $k$ neighboring points and $x^{seed}$. 
The specifics of the neighborhood search and the noisy interpolation subroutines of each method can vary. 
We detail the three methods studied in this paper hereafter.

\begin{algorithm}[h!]
  \caption{Seeded Local Combination (\SLC) Generation}
  \label{alg:sli_family}

  \SetKwFunction{NeighborhoodSearch}{NeighborhoodSearch}
  \SetKwFunction{NoisyInterp}{NoisyInterpolation}

  \Input{
    Dataset $D=\{x_1,\dots,x_n\}$, neighborhood size $k$,
    distance function $\mathrm{dist}(\cdot,\cdot)$.
  }
  \Output{Synthetic dataset $S$ of size $n$.}

  \BlankLine
  $\NeighborhoodSearch(D, x, k)$ returns the $k$ nearest neighbors of $x$ in $D$.
  Distances can be computed either in the input space or in a learned/projected (latent) space.

  \BlankLine
  $\NoisyInterp(x^{seed}, N)$ returns a synthetic sample obtained by a \emph{noisy combination} of (a subset of) obtained neighbors (and the seed). 
  The combination can be performed in the input space or in a latent space.

  \BlankLine
  $S \leftarrow \emptyset$\;

  \For{$r \leftarrow 1$ \KwTo $n$}{
    Select seed $x^{seed}$ from $D$\;

    $N \leftarrow \NeighborhoodSearch(D,\, x^{seed},\, k)$\;

    $x^{synth} \leftarrow \NoisyInterp(x^{seed},\, N)$\;

    $S \leftarrow S \cup \{x^{synth}\}$\;
  }

  \Return{$S$}\;

\end{algorithm}

While \SLC~generation does not directly modify the original profiles as anonymization techniques, it still relies on a selected seed profile to produce a synthetic observation. 
As noted previously~\cite{lebrun2024synthetic}, this relationship between original and synthetic data points potentially allows to conduct privacy attacks. 
As such \SLC~can be seen as an hybrid of synthetic data generation, which samples from the distribution, and anonymisation techniques, which modify profiles.

\emph{Synthetic Minority Oversampling Technique.}
SMOTE was originally proposed to address class imbalance issues in supervised learning~\cite{chawla2002smote} but can also be used to generate synthetic data. 
For each profile $x$ in $D$, SMOTE identifies its $k$ nearest neighbors and generates synthetic profiles by performing a randomized interpolation between the original instance and one of the identified neighbors.
SMOTE naturally fits in the \SLC~framework by instantiating the neighborhood search primitive as a standard search for closest neighbors. 
The noisy interpolation is performed by randomly selecting one neighbor and conducting a random linear interpolation between the seed point and the chosen neighbor. 
More formally, let $N$ be the set of the $k$ nearest neighbors of $x$ under the distance $\mathrm{dist}$.  
$x^{nn}$ is sampled from $N$ and $\lambda \sim \mathcal{U}(0,1)$, and then generate
\[
x^{synth} = x^{seed} + \lambda (x^{nn}-x^{seed}).
\]
\pagebreak

\emph{Avatar.} 
This method was proposed to generate synthetic profiles in the healthcare domain with ``no risk of re-identification''~\cite{guillaudeux2023patient} and also fits in the \SLC~framework.
The neighborhood search is performed in a latent space using a component analysis method (\emph{e.g.}, PCA), keeping only the first principal $m$ components. 
The noisy interpolation primitive is done in the latent space as a noisy weighted barycenter of the neighbors. 
More precisely, the weight used for the $i^{th}$ nearest neighbor when computing the barycenter is computed as:
$$W_i = \frac{D_i \times C_i \times R_i}{\sum_{i=1}^k D_i \times C_i \times R_i},$$
in which $D_i= \frac{1}{dist{(x^{seed},i)}}$,  with $dist(x^{seed},i)$ the distance between the profile of the seed and its $i^{th}$ neighbor, $C_i= \left(\frac{1}{2}\right)^j$ with $j$ randomly selected without replacement from the set $\{1,2,\ldots,k\}$ and $R_i \sim \mathrm{Exp}(1)$. 
Finally, the synthetic profile, called an avatar, is obtained by projecting back the obtained barycenter to the original data space.

\emph{Simulant.} 
This approach was also proposed as a synthetic data generation method in healthcare~\cite{beigi2023simulants}.
The neighborhood search primitive is similar to the one used in Avatar. 
However, the noisy interpolation is performed in the original data space with the synthetic profile being constructed feature by feature by randomly selecting a (set of) feature(s) from the $k$ neighbors of $x^{seed}$. 
A multiplicative Gaussian noise $\lambda \sim \mathcal{N}(1,\frac{1}{20})$ is then added to the neighbors feature value.
If some features are highly correlated, their values are sampled together from the same neighbor. 
While it is argued in the original paper that this addition of Gaussian noise ensures differential privacy, we challenge this claim in Section~\ref{sec:discussion}.

\section{Related work}
\label{sec:related}
In this section, we review three lines of work related to the privacy evaluation of synthetic data, focusing on the approaches most relevant to \SLC~generation. 
First, we describe post-hoc privacy metrics used in practice to evaluate the risks related to synthetic data releases before reviewing membership inference attacks against generative models. 
Finally, we discuss prior privacy evaluations of individual \SLC~ methods, in particular targeting SMOTE and Avatar.

\emph{Privacy metrics.}
Authors of the Avatar method~\cite{guillaudeux2023patient} as well as a following work using this method~\cite{demuth2025privacy} have assessed the privacy guarantees of synthetic data through the use of post-hoc metrics.
In their original work introducing Avatar the authors introduce four privacy metrics used to empirically measure residual privacy risks when releasing synthetic data~\cite{guillaudeux2023patient}. 
First, the distance-to-closest record (DCR) and the ratio between distance of closest and second closest neighbor (NNDR) are computed.
 For each synthetic record $s \in S$, its closest and second-closest distances to profiles in the original dataset $D$ can be defined as
 \[
 d_1(s,D) \;=\; \min_{x \in D} \mathrm{dist}(s,x),
 d_2(s,D) \;=\; \min_{x \in D \setminus \{x^\star(s)\}} \mathrm{dist}(s,x),
  \]
 in which
 \[
 x^\star(s) \;=\; \arg\min_{x \in D} \mathrm{dist}(s,x).
 \]
 The DCR and then the NNDR are computed as
 \[
 \mathrm{DCR}(s,D) \;=\; d_1(s,D),
 \qquad
 \mathrm{NNDR}(s,D) \;=\; \frac{d_1(s,D)}{d_2(s,D)}.
 \]
In addition to those metrics, they propose two additional privacy metrics specific to the Avatar method, namely the local cloaking and hidden rate metrics. 
The former counts for each real point $x_{seed}$ the number of synthetic profiles closer (in the latent space) to $x_{seed}$ than to the avatar derived from $x_{seed}$.
For each individual $x_i \in D$ producing a synthetic avatar $a_i$, with $dist_{lat}$ the Euclidean distance measured in latent space, the local cloaking is defined as 
\[
\mathrm{LC}(x_i)
=
\sum_{\substack{j=1\\ j\neq i}}^{n}
\mathbb{I}\!\left[dist_{lat}(x_i,a_j) < dist_{lat}(x_i,a_i)\right].
\]
Let $NN_S(x_i) = \arg\min_{a \in S} dist_{lat}(x_i,a)$ be the closest avatar to $x_i$.
The hidden rate measures the proportion of real profiles whose closest avatar (in the latent space) is not the avatar obtained from themselves (\emph{i.e.}, the percentage of observations with local cloaking whose value is not 0). 
More precisely, the hidden rate is defined as the proportion of individuals whose closest avatar is \emph{not} their own:
\[
\mathrm{HR}(D,S)
=
\frac{1}{n}\sum_{i=1}^{n}
\mathbb{I}\!\left[\mathrm{NN}_S(x_i) \neq a_i\right].
\]
The authors mention that those four metrics are related to the risk of singling out which according to the European Working party 29 \cite{WP2014} refers to the ability to identify in a dataset records linked to a particular individual.
However, we believe that they are also closely related to the risk on linkability. 
Indeed those metrics quantify how close the original profiles are to generated avatars, thus being correlated with how easy it is to link an original profile $x_i$ to its avatar $a_i$ obtained from seed $x_i$. 

More recently, a clinical study using the Avatar method relied on two additional post-hoc metrics included in the Avatar framework to asses privacy risks, namely the column direct match protection and row direct match protection~\cite{demuth2025privacy}. 
The column direct match protection refers to the maximum percentage of original profiles that cannot be linked to their produced avatar using the most informative column, or variable. 
The row direct match protection refers to the percentage of original profiles in which their produced avatar perfectly matches the original profile (\emph{i.e.}, data copying). 
Each post-hoc metric in the Avatar framework is associated to one of the three criteria defined by the Working Party 29~\cite{WP2014}.
More metrics can be found on 
the technical documentation of the commercial solution from Octopize~\cite{octopize_privacy_metrics}.
However, previous works have shown that 
DCR and other post-hoc distance-based privacy metrics despite being easy to interpret\footnote{A high distance between synthetic and real profiles can be interpreted as a sign that the synthetic data do not copy too closely the real data, thus being more private.} 
are poor indicators of the true privacy risks~\cite{ganev2025inadequacy,yao2025dcr}. 
Indeed, privacy attacks can still achieve high success rate even when distance-based metrics seem to indicate the contrary. 
One study has even shown that the per record DCR can be used to perform reconstruction attacks~\cite{ganev2025inadequacy}. 

\emph{Membership inference attacks.}
\label{related:MIA}
MIAs are a common approach to assess the privacy risks associated with a machine learning model the output it produces.  
In a MIA, the adversary aims at guessing if a specific profile was part of the training set of a machine learning model, which could be a generative model.  
Even if guessing membership might appear benign, it can be a serious privacy concern.
For instance, for the two datasets used in our experiments (AIDS and WBCD) being included in these datasets indicates respectively being infected with HIV or having breast tumor (benign or malignant).
Various methods for performing MIA on synthetic data generation have been developed recently.
Distance-based attacks such as GAN-leaks~\cite{chen2020gan} or Monte-Carlo membership~\cite{hilprecht2019monte} rely on the intuition that if a model has memorized a training sample it is more likely to produce synthetic profiles that are close to it. 
Another approach called DOMIAS~\cite{van2023membership} leverages auxiliary data to model the distributions of synthetic and auxiliary data.
Afterwards, the adversary computes the probability ratio of seeing observation $x$ coming from the synthetic versus from the auxiliary distributions. 
A high score indicates that point $x$ is much more likely under the synthetic distribution than the auxiliary distribution, and thus has a higher chance of being a member data point. 
More relevant to our work, Ward and collaborators~\cite{ward2025ensembling} have noticed that various MIAs do not identify the same profiles and suggest using ensemble approaches such as majority voting or weighted mean. 
We build on that intuition for designing our ensemble MIA. 

\emph{Privacy evaluation of \SLC }
Ganev and collaborators~\cite{ganev2025smotemirrorsexposingprivacy} have proposed a reconstruction attack called \textsf{ReconSmote}, which allows to reconstruct original data points from synthetic profiles generated by SMOTE~\cite{chawla2002smote}. 
This attack relies on the observation that SMOTE produces synthetic observation by interpolating between two real profiles. 
The attack works by identifying sets of three or more aligned synthetic profiles. 
Those three aligned synthetic data points should lie on a line in-between two real profiles. 
Thus, the intersection of two identified lines corresponds to a reconstructed real profile. 
In practice, \textsf{ReconSmote} achieves 100\% precision and an average of 85\% recall on eight different datasets. 
Note that this attack was performed in a data augmentation context, in which SMOTE was used to generate synthetic data points from a minority class. 
In this situation, real profiles are selected multiples times to produce synthetic profiles. 
We propose an extension of this attack to the more restrictive setting in which each real profile is only used once as seed to generate one synthetic profile and show how to apply this attack to synthetic data produced by Avatar with $k=2$.

Lebrun and collaborators have developed a distanced-based MIA tailored to Avatar, which measures distances in the latent space rather than in the original data space~\cite{lebrun2024synthetic}. 
In addition, they have also analyzed the privacy guarantees of the Avatar method by proposing a linkage attack.
Their linkage attack, which was called re-identification in their paper, aims at linking real profiles to synthetic profiles that are derived from them. 
Both attacks rely on distance-based metrics computed from the latent space formed by the principal component analysis.
In this work, we improve on their proposed approach for both the MIA and the linkage attack. 
More precisely, for the MIA, we consider their approach of measuring distances in the latent space as well as other distance-based proposed MIA from the literature and incorporate all attacks into an ensemble framework. 
For the linkage attack, we do not rely on distances between real and synthetic data points but rather on the likelihood of a synthetic data point coming from the distribution of synthetic points obtained from each real observation.

\section{Novel privacy attacks against \SLC}
\label{sec_attacks}

To assess the privacy guarantees of \SLC, we consider two well-known privacy goals~\cite{Salem2022SoKLT}, namely membership inference and reconstruction, and one legal anonymisation criteria from  the Article~29 Working Party (WP29) Opinion~05/2014 \cite{WP2014}, linkability.
In the case of synthetic data generation, the adversarial access to the model can range from a full access to model parameters (\emph{i.e.}, white-box access) to more restrictive settings in which the adversary has only query access to the attack model (\emph{i.e.}, black-box access) all the way to a regime in which the adversary only observes the synthetic data generated (\emph{i.e.}, no-box setting).
In this paper, we place ourselves in the challenging no-box setting, considering that the adversary is aware only of the specific \SLC~ algorithm used and the choice of the values of its hyper-parameters (\emph{i.e.}, the adversary can run the \SLC~algorithm by himself). 
In addition, we also present in Appendix A 
results from an Attribute Inference Attack (AIA). 
This attack naturally fits within the inference legal anonymization criterion from the Article~29 Working Party (WP29) Opinion~05/2014~\cite{WP2014} making it relevant to evaluate privacy leakage in \SLC. 

\subsection{Membership inference attack}
\label{sec:MIA}

The success of an MIA can be evaluated by first sampling a training dataset $D_{train}$ of size $n$ from private dataset $D$.
Then, a \SLC~synthetic data generation algorithm is applied on $D_{train}$ to produce the synthetic dataset $S$ of size $n$. 
An equal number $m$ of members and non-members are sampled respectively from $D_{train}$ and $D_{non\_member} = D \setminus D_{train}$ to obtain the guessing set $D_{guess}$. 
The attacker objective is to predict for each profile from $D_{guess}$ if it belongs to $D_{member}$ or $D_{non\_member}$.
When evaluating the privacy risk using an MIA, a small number of correct confident guesses is usually considered a higher privacy risk than a larger number of non-precise guesses.
Following this principle, we will present the MIA results as suggested by Carlini and collaborators~\cite{carlini2022membership} and widely adopted in the literature using True Positive Rate (TPR) at low False Positive Rate (FPR) curve and reporting the TPR@ 10\% FPR.

One of our contribution is the design of an ensemble approach for MIA.
In this approach, we first perform the individual attacks previously described in Section~\ref{related:MIA}, computing each distanced-based attack score both in the original data space as well as in the latent space.
Scores from individual attacks concatenated with the original features are then passed through a Gradient Boosting model~\cite{friedman2001greedy} to produce the final membership score. 
Note that the original feature vector is passed as input to the attack model is because some attacks are more efficient in various parts of the data space (\emph{e.g.}, having a close neighbor does not mean the same thing in a dense part or at the edge of the distribution).
We apply shadow modeling with 50 shadow models to build the attack dataset used to train the Gradient Boosting model.
Algorithm~\ref{alg:ensemble_mia} describes the attack pipeline.

\begin{algorithm}[h!]
  \caption{Ensemble MIA attack}
  \label{alg:ensemble_mia}

  \SetKwFunction{Gen}{SLC\_Generate}
  \SetKwFunction{Sample}{Sample}
  \SetKwFunction{AttackScore}{AttackScore}
  \SetKwFunction{Feat}{BuildAttackFeatures}
  \SetKwFunction{TrainGB}{TrainGB}
  \SetKwFunction{PredictGB}{PredictGB}

  \Input{
    Guessing dataset $D_{guess}$, Synthetic dataset $S$ obtained from $D_{guess}$ ,number of shadow iterations $t$, shadow dataset size $m$,
    set of base attacks $\mathcal{A}=\{A_1,\dots,A_L\}$ each providing a score, 
    supervised learner $\mathrm{GB}$ (Gradient Boosting).
  }
  \Output{
    Final membership scores $\{\hat{a}(x) : x \in D_{\mathrm{guess}}\}$.
  }

  \BlankLine
  $\Gen(\cdot)$ producing synthetic records;
  
  $\Feat(x, S, \mathcal{A})$ returns a feature vector built from (1) base-attack scores computed in data space and/or latent space and (2) the original feature vector of $x$.
  
  \BlankLine
  \textbf{Shadow modeling (\emph{i.e.}, attack dataset construction).}\\
  $D_{\mathrm{shadow}} \leftarrow \emptyset$\;

  \For{$i \leftarrow 1$ \KwTo $t$}{
    $D_{\mathrm{shadow\_member}} \leftarrow \Sample(D_{\mathrm{guess}},\, m)$\;
    $D_{\mathrm{shadow\_nonmember}} \leftarrow \Sample(D_{\mathrm{guess}}\setminus D_{\mathrm{shadow\_member}},\, m)$\;

    $S_{\mathrm{shadow}} \leftarrow \Gen(D_{\mathrm{shadow\_member}})$\;

    \ForEach{$x \in D_{\mathrm{shadow\_member}} \cup D_{\mathrm{shadow\_nonmember}}$}{
      $y(x) \leftarrow \mathbb{I}[x \in D_{\mathrm{shadow\_member}}]$\;
      $v(x) \leftarrow \Feat(x, S_{\mathrm{shadow}}, \mathcal{A})$\;
      $D_{\mathrm{shadow}} \leftarrow D_{\mathrm{shadow}} \cup \{(v(x), y(x))\}$\;
    }
  }

  \BlankLine
  \textbf{Training of the ensemble model.}\\
  $\mathrm{GB} \leftarrow \TrainGB(D_{\mathrm{shadow}})$\;

  \BlankLine
  \textbf{Final prediction on the guessing set.}\\
  \ForEach{$x \in D_{\mathrm{guess}}$}{
    $v(x) \leftarrow \Feat(x, S, \mathcal{A})$\;
    $\hat{a}(x) \leftarrow \PredictGB(\mathrm{GB}, v(x))$\;
  }

  \Return{$\{\hat{a}(x) : x \in D_{\mathrm{guess}}\}$}\;
\end{algorithm}

\subsection{Likelihood-based linkage attack}
\label{sec:linkage}
 
In this section, we propose a linkage attack, in which given the original dataset and the synthetic dataset the adversary tries to link back the synthetic profile to its original profile (\emph{i.e.}, identifying an individual corresponding to a synthetic profile).
The adversarial setting hardly corresponds to a real-life context due to the inclusion of the full original dataset in the adversarial background knowledge. 
However, this linkage attack can be used for benchmarking purposes. 
Furthermore, the attack closely aligns with the definition of anonymized data set by the GDPR~\cite{gdpr} making it an useful tool for practitioners to assess the privacy of synthetic data. 
The first step of the attack consists in generating a large number of synthetic profiles for each of the original profiles, and then to estimate a distribution from these synthetic profiles.

\begin{figure}[h!]
        \centering
        \vspace{-3mm}
        \includegraphics[width=0.9\linewidth]{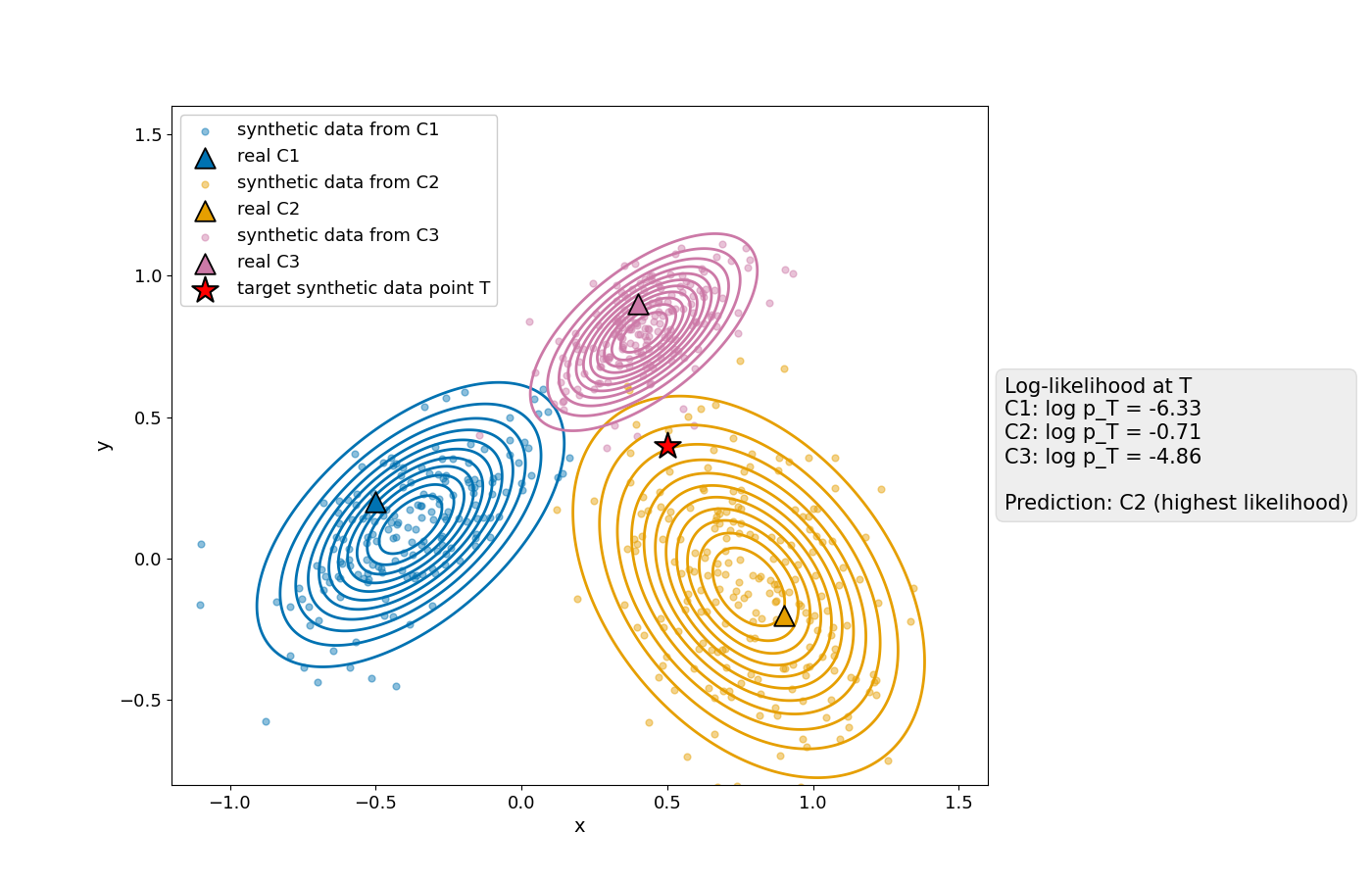}
        \caption{Illustration of the linkage attack using multivariate normal distributions.}
        \label{fig:linkage_attack}
\end{figure}

Figure~\ref{fig:linkage_attack} illustrates the attack in this context. 
The colored points represent avatars generated from three different original profiles and their respective estimated multivariate densities. 
In this example, while the target synthetic profile is closest in distance to the profile $C3$, the densities indicate that it is most likely to have been generated by the profile $C2$, which illustrates how linking simulated profiles to the closest real observation can be sub-optimal. 
Algorithm~\ref{alg:linkage} presents the overall approach for the linkage attack. 
First, we run $t=200$ simulations, producing a new synthetic data point from real data each time. 
Then, for each original profile $x_{seed}$ we model the distribution of synthetic data points produced from this seed. 
The implementation of the \textsf{ModelDistribution} primitive varies depending on the \SLC~algorithm. 
For Avatar, we estimate the parameters $\mu_{seed}$ and $\Sigma_{seed}$ of multivariate normal distributions. 
For each candidate $x_{seed}$, we model the distribution of its avatars as
$$
s \sim \mathcal{N}\!\left(\mu_{seed}, \Sigma_{seed}\right).
$$
Then for a target avatar $s$, the log-likelihood of observing it under the distribution of avatars obtained from $x_{seed}$ is computed as:
\[
\begin{aligned}
 {}&\log p\!\left(s \mid \mu_{\mathrm{seed}}, \Sigma_{\mathrm{seed}}\right)
= \\
& -\frac{1}{2}\Big[
\log \big|\Sigma_{\mathrm{seed}}\big| 
+ (s-\mu_{\mathrm{seed}})^\top
\Sigma_{\mathrm{seed}}^{-1}
(s-\mu_{\mathrm{seed}})
+ d\log(2\pi)
\Big].
\end{aligned}
\]
In practice, $\Sigma_{seed}$ may not be invertible as a singular covariance matrix due to the data lying in a lower dimension sub-space or features with (nearly) constant value leading to (almost) null variance.
This can be handled in practice by adding a small ridge correction,  projecting the data in a lower dimensional space with PCA and/or relying on the global covariance estimate obtained across all avatars simulated for all profiles.
For SMOTE and Simulant, we fit a Gaussian Kernel Density Estimator (KDE) as for these methods the distributions of simulated synthetic profiles for a given seed $x_{seed}$ are not-well modeled by multivariate normal distributions. 
Then, for each synthetic profile $s$, we identify its $k'$ nearest real profiles, which we consider as potential seeds.
For each of these $k'$ potential seeds, we compute the log-likelihood of observing synthetic profile $s$ given the estimated distribution of points originated from $x_{seed}$. 
Finally, we compute the likelihood ratio as a measure of confidence of the association predicted. 
We use a top-1 (highest) vs top-2 (second highest) log-likelihood to quantify how much more plausible the best candidate seed is compared to the runner-up.

\begin{algorithm}[h!]
  \caption{Linkage attack against \SLC~generators via per-seed likelihood modeling}
  \label{alg:linkage}

  \SetKwFunction{Gen}{SLC\_Generate}
  \SetKwFunction{ModelDist}{ModelDistribution}
  \SetKwFunction{Lik}{Log-Likelihood}
  \SetKwFunction{KNN}{kNN}

  \Input{
    Real dataset $D=\{x_1,\dots,x_n\}$, released synthetic dataset $S$,
    \SLC generator $\Gen(\cdot)$, number of simulations $t$,
    candidate set size $k'$.
  }
  \Output{
    Link scores $\{\mathrm{score}(s) : s \in S\}$ and predicted links $x^\star(s)$.
  }

  \BlankLine
  $\ModelDist(\{s^{(1)},\dots,s^{(t)}\})$ fits a parametric or non-parametric model for the seed-conditional
  distribution of synthetic outputs, which for Avatar corresponds to estimate $(\mu_{\mathrm{seed}}, \Sigma_{\mathrm{seed}})$ while for SMOTE/Simulant it refers to fitting a Gaussian KDE.

  \BlankLine
  $\Lik(s \mid \mathcal{M}_{\mathrm{seed}})$ returns the log-likelihood (or density) of $s$ under the fitted model
  $\mathcal{M}_{\mathrm{seed}}$.

  \BlankLine
  \textbf{Step 1: build per-seed synthetic-output distributions.}\\
  \For{$r \leftarrow 1$ \KwTo $t$}{
    $S^{(r)} \leftarrow \Gen(D)$\;
  }

  \ForEach{$x_{\mathrm{seed}} \in D$}{
    Collect $\mathcal{S}_{\mathrm{seed}} \leftarrow \{\, s^{(r)}_{\mathrm{seed}} : r \in \{1,\dots,t\}\,\}$\;
    $\mathcal{M}_{\mathrm{seed}} \leftarrow \ModelDist(\mathcal{S}_{\mathrm{seed}})$\;
  }

  \BlankLine
  \textbf{Step 2: score candidate links for each released synthetic record.}\\
  \ForEach{$s \in S$}{
    $C(s) \leftarrow \KNN(D, s, k')$\;

    \ForEach{$x_{\mathrm{seed}} \in C(s)$}{
      $\ell(s \mid x_{\mathrm{seed}}) \leftarrow \Lik(s \mid \mathcal{M}_{\mathrm{seed}})$\;
    }

    Let $x^\star(s) \leftarrow \arg\max_{x_{\mathrm{seed}} \in C(s)} \ell(s \mid x_{\mathrm{seed}})$\;
    Let $x^{(2)}(s)$ be the second-best seed in $C(s)$ according to $\ell(s \mid x_{\mathrm{seed}})$\;

    $\mathrm{score}(s) \leftarrow
       \ell(s \mid x^\star(s)) -  \ell(s \mid x^{(2)}(s))$\;
  }

  \Return{$\{x^\star(s), \mathrm{score}(s) : s \in S\}$}\;
\end{algorithm}

To assess the performance of the general linkage attack, we consider three baselines. 
The first one assumes that the data is truly anonymous if the adversary succeeds at linking a synthetic profile with the seed it originates from no better than with a random guess, which has success probability of  $\frac{1}{n}$.
We also consider another baseline in which each avatar is linked to the closest real profile.
Finally, we compare ourselves to the linkage attack proposed by Lebrun and collaborators, which is similar to the second baseline but performed in a PCA latent space fitted by the adversary on real data~\cite{lebrun2024synthetic}. 
In addition to this generic scenario, which describes a privacy evaluation tailored to the definition of anonymous data, we provide in Appendix C 
a more realistic attack setting in which the adversary only has access to the synthetic dataset as well as one randomly sampled real profile $x_{seed}$. 
In this context, the main objective of the adversary is to identify in the synthetic dataset the synthetic observation obtained from $x_{seed}$. 

\subsection{Reconstruction attack}

Finally, we extend a reconstruction attack design for SMOTE (called \textsf{ReconSmote})~\cite{ganev2025smotemirrorsexposingprivacy} to the Avatar in the specific case in which $k=2$. 
This setting has been used previously in practice for the open-data release of two medical datasets~\cite{demuth2025privacy,paris2025genetic}. 
Recall that the Avatar method with $k=2$ interpolates between the two closest neighbors of the seed point, so that synthetic profiles are guaranteed to be in between two real points. 
In our case, we do not assume that we are oversampling any minority class, nor that we generate more synthetic data points than the number of original data points. 
As a consequence, selectively repeating (\emph{i.e.}, 3 times or more as in \textsf{ReconSMOTE}) the same pair of observations as closest neighbors from different seeds is unlikely.

Our proposed extension of the \textsf{ReconSMOTE} attack, in addition to finding sets of three aligned data points, searches for pairs of synthetic profiles close to each other (likely to be between the same two real profiles) and we look at intersections (\textit{i.e.}, reconstructed profiles) that are within a close radius of previously identified synthetic data points. 
Each line is assumed to correspond to the segment between two real endpoints, thus intersections of such lines are candidate reconstructions of real records as in \textsf{ReconSMOTE}.
Algorithm~\ref{alg:recon_avatar_k2} describes in details the proposed reconstruction attack.
Since our approach relaxes the need to find three aligned synthetic points and instead mainly relies on pairs of close synthetic points, our approach will likely not achieve the same 100\% precision of \textsf{ReconSMOTE}.
Therefore, to increase the precision of prediction, a simple filtering rule was implemented. 
More precisely if a given point is reconstructed multiple times, this indicates multiple segments supporting pairs of close avatar pointing to this point (\emph{i.e.}, the same point is being used in the construction of multiple avatars). 
Thus, the reconstructed points are filtered to keep only points that are most often reconstructed.

\begin{algorithm}[h!]
  \caption{Reconstruction attack for Avatar with $k=2$}
  \label{alg:recon_avatar_k2}

  \SetKwFunction{Triplets}{FindColinearTriplets}
  \SetKwFunction{Pairs}{FindClosePairs}
  \SetKwFunction{FitLine}{FitLine}
  \SetKwFunction{Dedup}{DeduplicateLines}
  \SetKwFunction{Intersections}{LineIntersections}
  \SetKwFunction{Near}{HasNearbyPoint}
  \SetKwFunction{Dist}{dist}

  \Input{
    Synthetic dataset $S=\{s_1,\dots,s_N\}$ generated by Avatar with $k=2$,
    colinearity tolerance $\varepsilon$,
    neighborhood thresholds $\tau_1$ and $\tau_2$,
    distance function $\Dist(\cdot,\cdot)$.
  }
  \Output{Set of reconstructed candidate real points $\widehat{D}$.}

  \BlankLine
  $\Triplets(S,\varepsilon)$ returns triplets $(s_a,s_b,s_c)$ that are colinear up to tolerance $\varepsilon$.

  \BlankLine
  $\Pairs(S,\tau_1)$ returns pairs $(s_u,s_v)$ such that $\Dist(s_u,s_v)\le \tau_1$.

  \BlankLine
  $\FitLine(\{s_1,\dots,s_m\})$ returns a line $\ell$ supporting the points.

  \BlankLine
  $\Near(\ell, P, \tau_2)$ returns \texttt{true} if $\min_{s\in P}\Dist(s,\ell)\le \tau_2$.

  \BlankLine
  $L \leftarrow \emptyset$\;

  \BlankLine
  \textbf{Step 1: find lines from close colinear triplets.}\\
  \ForEach{$(s_a,s_b,s_c)\in \Triplets(S,\varepsilon)$}{
    \If{$\max\{\Dist(s_a,s_b),\Dist(s_a,s_c),\Dist(s_b,s_c)\}\le \tau_1$}{
      $\ell \leftarrow \FitLine(\{s_a,s_b,s_c\})$\;
      $P_l \leftarrow \{s_a,s_b,s_c\}$\;
      $L \leftarrow L \cup \{\ell\}$\;
    }
  }

  \BlankLine
  \textbf{Step 2: lines from close pairs.}\\
  \ForEach{$(s_a,s_b)\in \Pairs(S,\tau_1)$}{
    $\ell \leftarrow \FitLine(\{s_a,s_b\})$\;
    $P_l \leftarrow \{s_a,s_b\}$\;
    $L \leftarrow L \cup \{\ell\}$\;
  }

  \BlankLine
  \textbf{Step 3: de-duplicate lines.}\\
  $L \leftarrow \Dedup(L)$\;

  \BlankLine
  \textbf{Step 4: intersect lines and keep plausible reconstructions.}\\
  $\widehat{D} \leftarrow \emptyset$\;
  $\mathcal{I} \leftarrow \Intersections(L)$\;

  \ForEach{$(i,\ell_1,\ell_2)\in \mathcal{I}$}{
    \If{$\Near(i,P_{l_1},\tau_2)$ \textbf{and} $\Near(i,P_{l_2},\tau_2)$}{
      $\widehat{D} \leftarrow \widehat{D} \cup \{i\}$\;
    }
  }

  \Return{$\widehat{D}$}\;
\end{algorithm}

\section{Experimental evaluation}
\label{sec:exp_evaluation}

Hereafter, we detail the experiments that we have conducted to assess the privacy guarantees provided by \SLC~approaches through the lens of the privacy attacks introduced in the previous section.
We do not report on the exact computational times for our attacks since they all run under one hour on a laptop with 16 CPU cores and 32 GB RAM. 

\subsection{Experimental setting}

The performance of the attacks has been evaluated on three classical datasets. 
The first two datasets are the Wisconsin Breast Cancer Dataset WBCD~\cite{wolberg1992breastcancerwi} and AIDS~\cite{aids_uci}, which have been widely used including in the paper proposing the Avatar approach. 
The WBCD dataset contains 9 ordinal categorical columns and one binary target with 449 profiles, with the associated prediction task being to predict whether a tumor is benign or malignant. 
The AIDS dataset contains a mix of 26 categorical and continuous features with 2139 profiles, with the classification task being to predict whether a patient affected by HIV has reached the AIDS phase or not. 
Finally, we also use the California Housing dataset~\cite{pace1997sparse}, which consists of 20,640 profiles with 9 continuous features. 
The regression task consists of predicting the sale price of Californian properties.
Together, these three datasets provide a good coverage in terms of dataset size and feature types.  

As a preprocessing step, we remove from each dataset all the duplicates as they could lead to the generation of duplicates between synthetic and real data as we want to avoid such direct violation of privacy in our evaluation. 
In our experiments, the number of principal components kept for the avatar generation is half the number of dimensions in the original dataset. 
We present in Appendix F
additional results in which we have varied the number of components kept in the latent space. 
The parameter $k$ is the main privacy parameter in each of the considered \SLC~methods, as it controls how many neighbors are used to produce the synthetic data point from the seed. 
Intuitively, for Simulant and Avatar, bigger values of $k$ dilute the importance that each neighbor takes in the computation of the synthetic data point.
Previous work report using values  of $k \in \{2,5,10,20\}$ with $5$ and $10$ being default values respectively for Simulant and Avatar. 
For instance for the Avatar method, $k=2$ and $k=5$ are the values being reported in previous real-world studies having used Avatar~\cite{demuth2025privacy, paris2025genetic}, $k=10$ is the default value and $k=20$ is the value used in the original avatar paper~\cite{guillaudeux2023patient} as well as in the study by Lebrun and collaborators~\cite{lebrun2024synthetic}.

\begin{table}[h!]
\centering
\small
\setlength{\tabcolsep}{5pt}
\renewcommand{\arraystretch}{1.08}
\caption{Accuracy comparison between raw data and \SLC-generated synthetic data for different neighborhood sizes $k$.}
\begin{tabular}{l l c c c}
\hline
\textbf{Method} & \textbf{$k$} & \textbf{WDBC (acc)} & \textbf{AIDS (acc)} & \textbf{\makecell{California\\Housing (MSE)}} \\
\hline
\multicolumn{2}{l}{\textbf{Raw data}} & 97.78\% & 90.19\% & 0.1950 \\
\hline
\multirow{4}{*}{\textbf{Avatar}}   & 2  & 93.33\% & 85.75\% & 0.2337 \\
                                  & 5  & 95.56\% & 81.31\% & 0.2726 \\
                                  & 10 & 87.78\% & 79.44\% & 0.2650 \\
                                  & 20 & 77.78\% & 76.40\% & 0.2742 \\
\hline
\multirow{4}{*}{\textbf{SMOTE}}    & 2  & 97.78\% & 89.25\% & 0.2339 \\
                                  & 5  & 95.56\% & 89.49\% & 0.2287 \\
                                  & 10 & 95.56\% & 89.02\% & 0.2353 \\
                                  & 20 & 98.89\% & 88.79\% & 0.2352 \\
\hline
\multirow{4}{*}{\textbf{Simulant}} & 2  & 95.56\% & 87.15\% & 0.3272 \\
                                  & 5  & 94.44\% & 83.64\% & 0.5125 \\
                                  & 10 & 94.44\% & 84.81\% & 0.5770 \\
                                  & 20 & 96.67\% & 80.84\% & 0.6228 \\
\hline
\end{tabular}
\label{tab:utility}
\end{table}

Table~\ref{tab:utility} presents the utility in term of ML accuracy for each \SLC~generative method as well as baseline performance using real data for the three datasets for various values of $k$. 
The machine learning accuracy is measured on a real hold-out test set with a XGBoost model trained on synthetic dataset (or real data for the baseline) with standard hyper parameter tuning (number of tree, depth, learning rate, \ldots) performed for 100 steps. 

For both AIDS and WDBC dataset, the utility of the baseline is quite high and coherent with expected values for those datasets ~\cite{wolberg1992breastcancerwi, aids}.
For both classification tasks, we can observed that models trained on synthetic data produced by SMOTE slightly outperformed models trained on synthetic data points from Simulant and Avatar. 
In addition, the machine learning efficiency decreases for those two methods when $k$ increases while SMOTE does not suffer to a decrease in utility.

Table~\ref{tab:cloaking_hidden_rate} describes the privacy evaluation with post-hoc metrics as proposed by Avatar. 
We have slightly adapted the metrics for SMOTE by measuring them in the data space rather than in the latent space.
Since this generative method does not make use of a latent space, we believe that staying in the original data space is more relevant. 

Simulant seems to be the \SLC~offering the better privacy guarantees according to the selected metrics. 
Except for extremely low values for $k$, the hidden rate is mostly above 95\% while linkage attacks based on distance only would mostly fail.
Synthetic profiles produced by SMOTE seem to be highly vulnerable to distance-based attacks as for half of the synthetic data the closest real observation is their real counterpart.  

\begin{table}[h!]
\centering
\small
\setlength{\tabcolsep}{5pt}
\renewcommand{\arraystretch}{1.08}
\caption{Privacy metrics by dataset, \SLC~method and neighborhood size $k$. The hidden rate is reported in percentage while the local cloaking corresponds to the median cloaking value.}
\label{tab:cloaking_hidden_rate}
\resizebox{\columnwidth}{!}{%
\begin{tabular}{llcccccc}
\toprule
\multirow{2}{*}{Dataset} & \multirow{2}{*}{$k$}
& \multicolumn{2}{c}{\textbf{Avatar}}
& \multicolumn{2}{c}{\textbf{SMOTE}}
& \multicolumn{2}{c}{\textbf{Simulant}} \\
\cmidrule(lr){3-4}\cmidrule(lr){5-6}\cmidrule(lr){7-8}
& & Hidden rate & Local cloaking & Hidden rate & Local cloaking & Hidden rate & Local cloaking \\
\midrule
\multirow{4}{*}{WBCD}
& 2  & 97.77\% & 69.0   & 38.98\% & 0.0 & 67.71\% & 2.0    \\
& 5  & 98.22\% & 78.0   & 38.98\% & 0.0 & 90.87\% & 43.0   \\
& 10 & 97.55\% & 85.0   & 46.10\% & 0.0 & 95.55\% & 113.0  \\
& 20 & 98.00\% & 91.0   & 50.11\% & 0.0 & 98.89\% & 134.0  \\
\midrule
\multirow{4}{*}{AIDS}
& 2  & 86.07\% & 2.0    & 34.92\% & 0.0 & 83.08\% & 11.0   \\
& 5  & 84.39\% & 2.0    & 36.75\% & 0.0 & 97.05\% & 103.0  \\
& 10 & 86.86\% & 3.0    & 41.84\% & 0.0 & 98.41\% & 184.0  \\
& 20 & 91.77\% & 5.0    & 44.46\% & 0.0 & 99.53\% & 228.0  \\
\midrule
\multirow{4}{*}{\makecell{California\\Housing}}
& 2  & 93.77\% & 11.0   & 37.11\% & 0.0 & 87.72\% & 108.0  \\
& 5  & 91.16\% & 9.0    & 41.52\% & 0.0 & 97.86\% & 679.5  \\
& 10 & 91.93\% & 12.0   & 45.42\% & 0.0 & 99.07\% & 968.0  \\
& 20 & 93.20\% & 16.0   & 49.53\% & 0.0 & 99.55\% & 1174.5 \\
\bottomrule
\end{tabular}%
}
\end{table}

\subsection{MIA}

\begin{figure*}[h!]
    \centering
    \begin{subfigure}{0.32\textwidth}
        \centering
        \includegraphics[width=\linewidth]{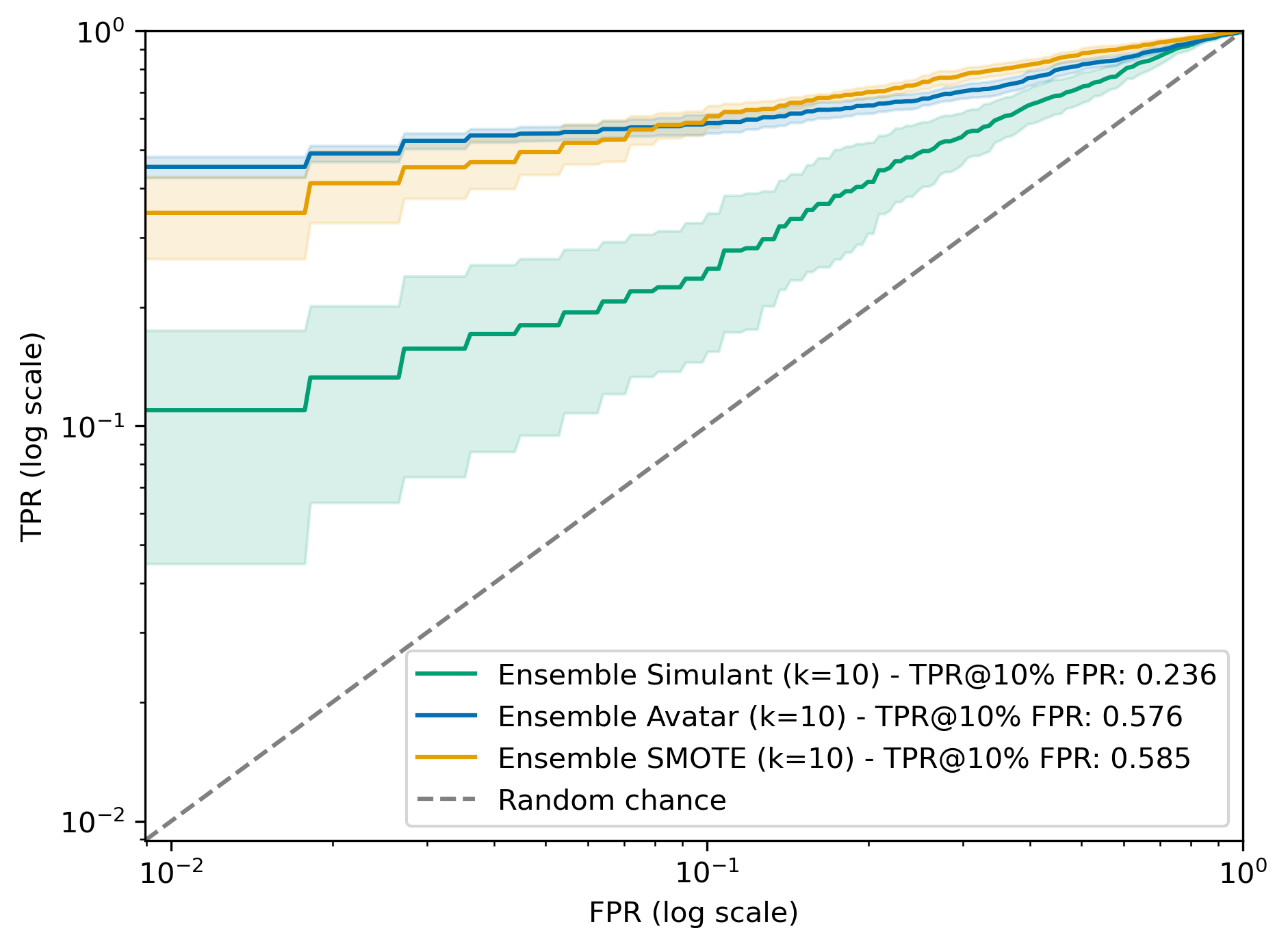}
        \caption{WBCD}
        \label{fig:MIA_breast}
    \end{subfigure}
    \hfill
    \begin{subfigure}{0.32\textwidth}
        \centering
        \includegraphics[width=\linewidth]{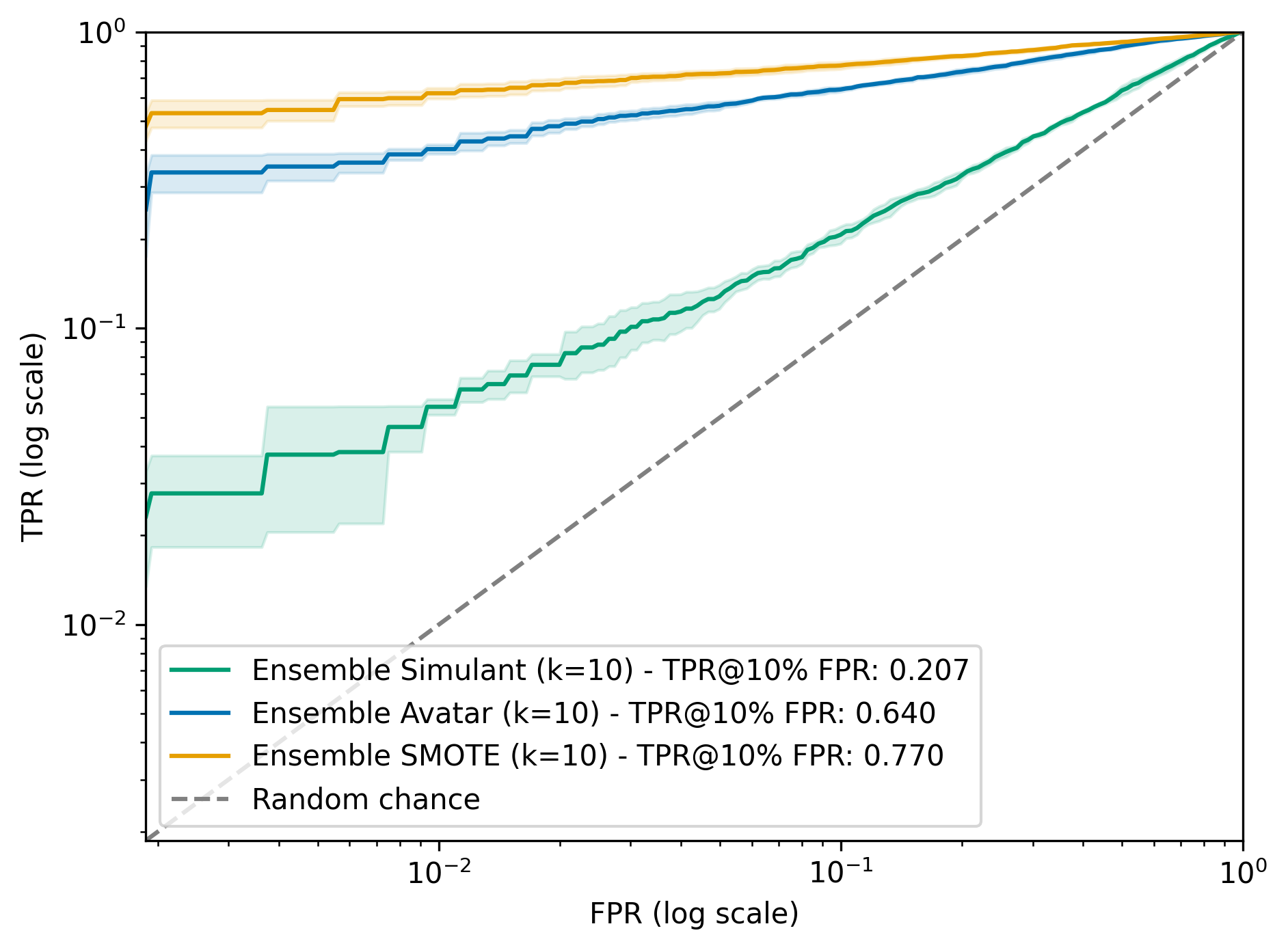}
        \caption{AIDS}
        \label{fig:MIA_aids}
    \end{subfigure}
    \hfill
    \begin{subfigure}{0.32\textwidth}
        \centering
        \includegraphics[width=\linewidth]{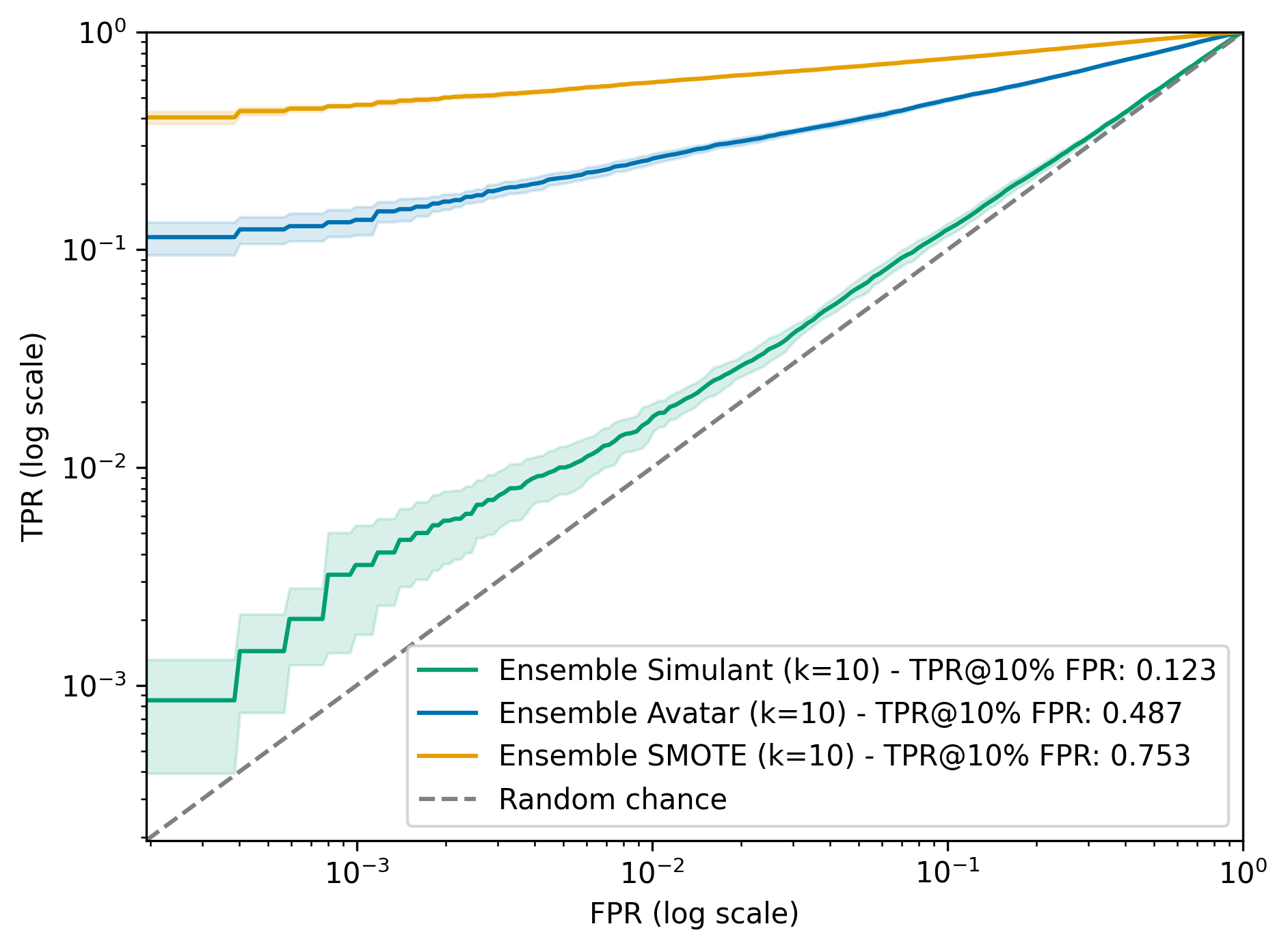}
        \caption{California Housing}
        \label{fig:MIA_california}
    \end{subfigure}

    \caption{Ensemble MIA success rate against Avatar, Simulant and SMOTE (for $k=10$).}
    \label{fig:MIA_three_datasets}
\end{figure*}

Figure~\ref{fig:MIA_three_datasets} presents the results of MIAs performed against each \SLC~method using $k=10$ for each dataset.
Regardless of the dataset, we observe the same trend, which is that synthetic profiles produced by Simulant are less vulnerable to MIA compared to SMOTE and Avatar. 
Both SMOTE and Avatar are extremely vulnerable to our ensemble MIA with TPR\@10\%FPR being between nearly five times to seven times better than random guessing.
This attack success rate in MIA against synthetic tabular data generation is critically high, and has been rarely observed in the MIA literature.
This highlights a rather worrying privacy leakage.
While ensemble MIA is less successful on the WBCD and AIDS datasets for Simulant, a TPR\@10\%FPR approximately equal to 20\% still indicates a serious privacy leakage. 

\begin{figure*}[h!]
    \centering
    \begin{subfigure}{0.32\textwidth}
        \centering
        \includegraphics[width=\linewidth]{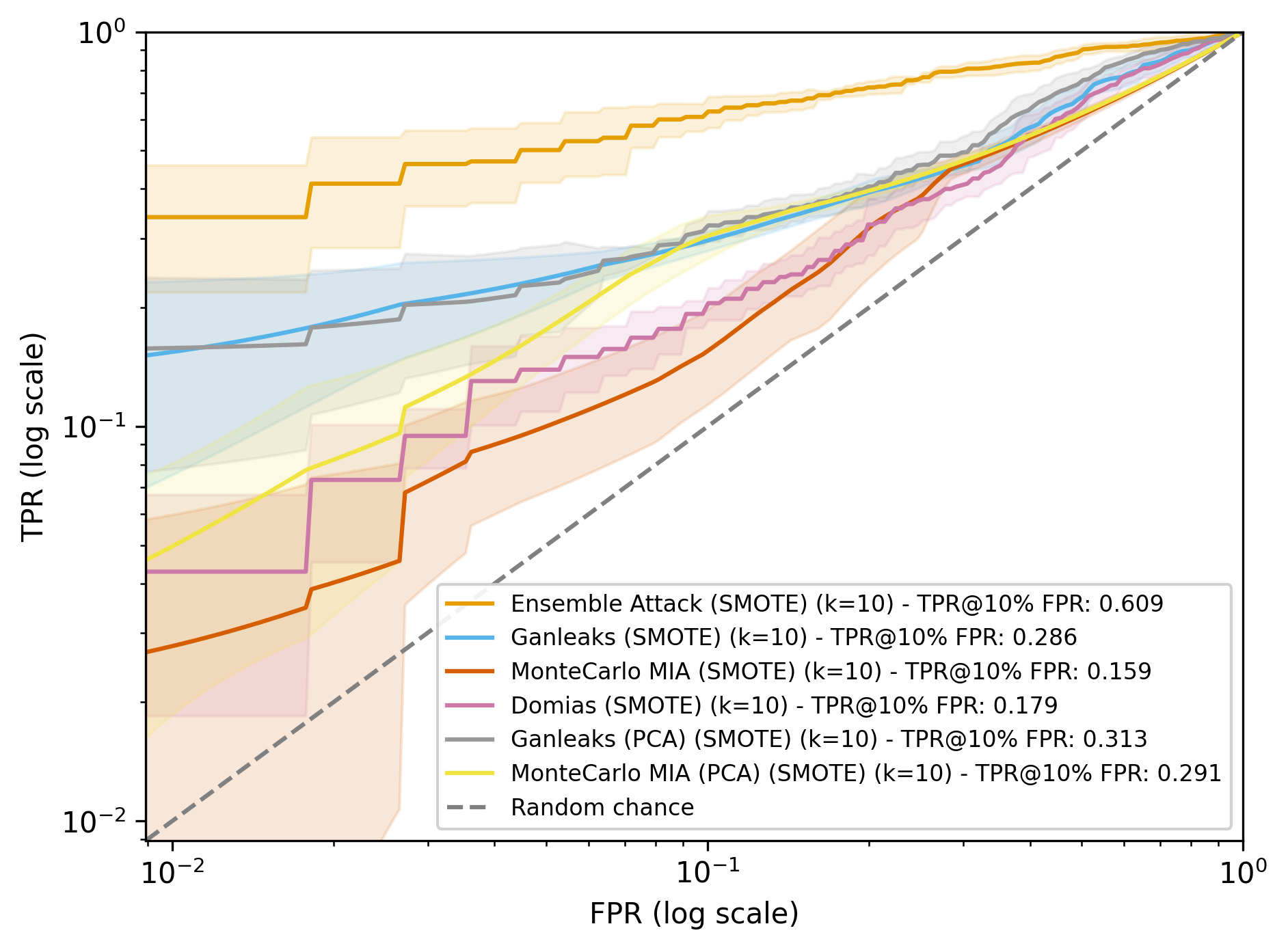}
        \caption{WBCD}
        \label{fig:MIA_ensemble_breast}
    \end{subfigure}
    \hfill
    \begin{subfigure}{0.32\textwidth}
        \centering
        \includegraphics[width=\linewidth]{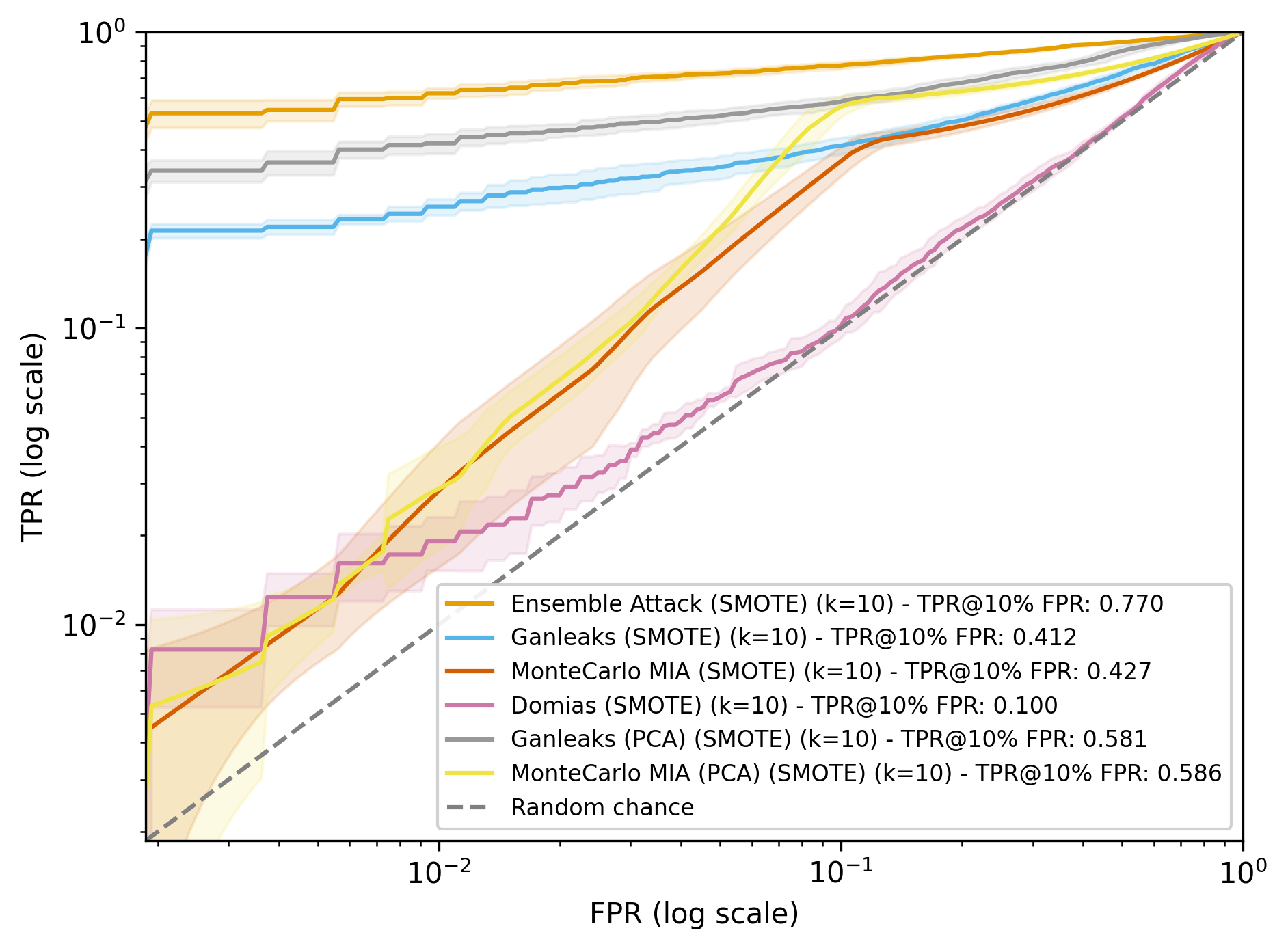}
        \caption{AIDS}
        \label{fig:MIA_ensemble_aids}
    \end{subfigure}
    \hfill
    \begin{subfigure}{0.32\textwidth}
        \centering
        \includegraphics[width=\linewidth]{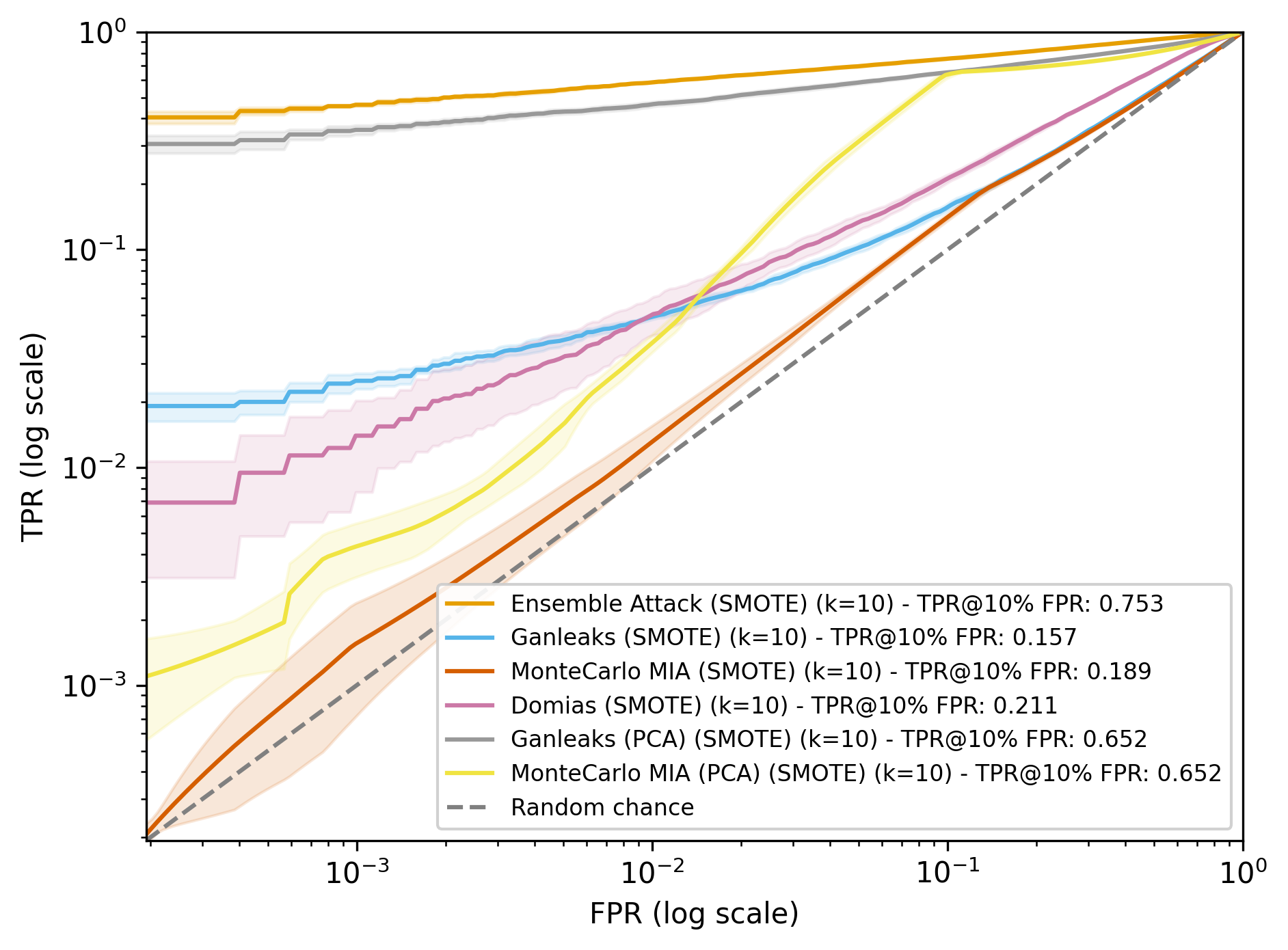}
        \caption{California Housing}
        \label{fig:MIA_ensemble_california}
    \end{subfigure}

    \caption{Ensemble MIA vs individual attacks for synthetic data produced with SMOTE ($k=10$).}
    \label{fig:MIA_ensemble}
\end{figure*}

Figure~\ref{fig:MIA_ensemble} illustrates how the ensemble MIA performs better than each individual MIA across all datasets.
We can observe that for the same generative method, individual attacks do not perform equally well across datasets and generative method with no single MIA outperforming the others.

\begin{figure*}[h!]
    \centering
    \begin{subfigure}{0.32\textwidth}
        \centering
        \includegraphics[width=\linewidth]{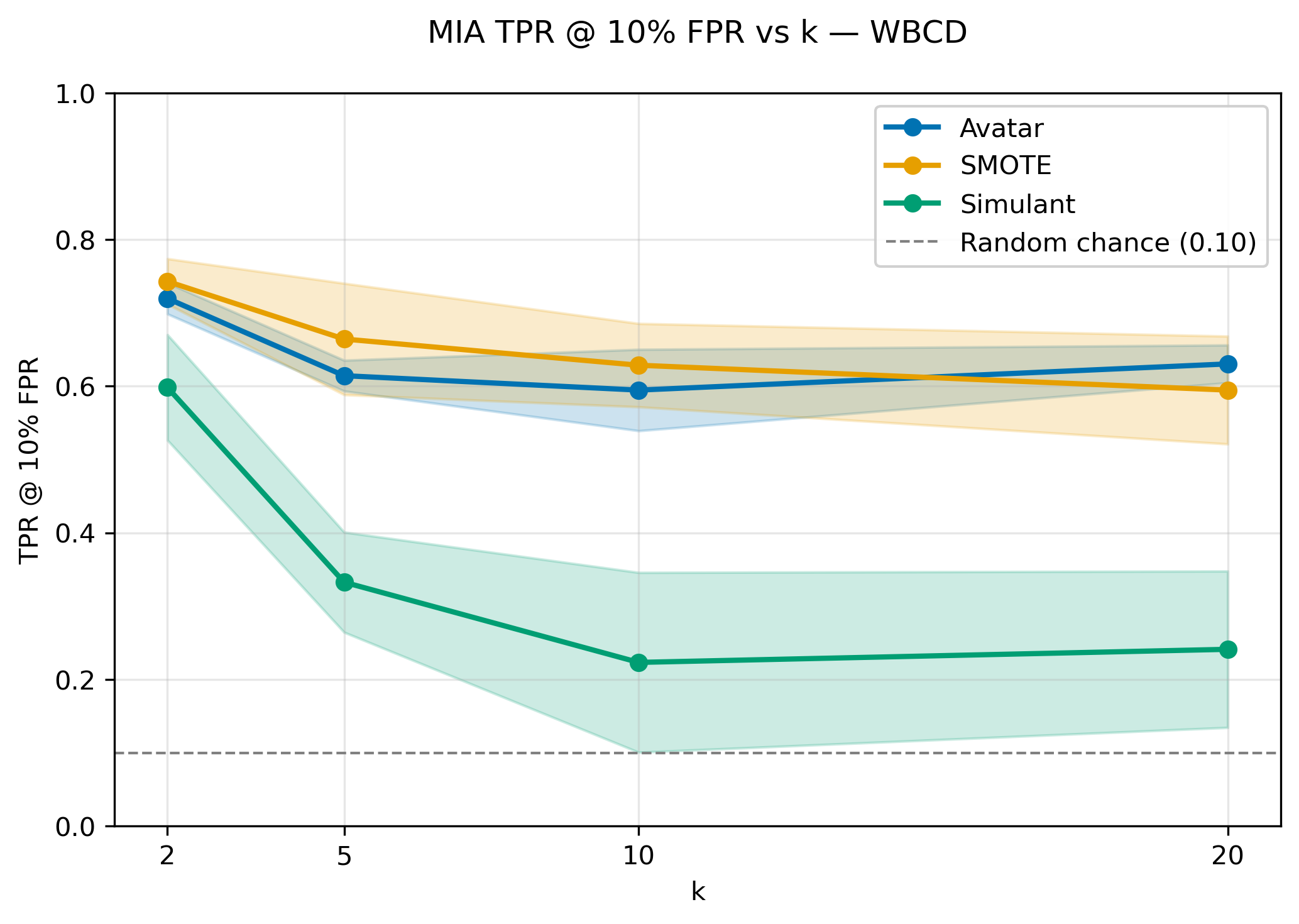}
        \caption{WBCD}
        \label{fig:linkage_breast}
    \end{subfigure}
    \hfill
    \begin{subfigure}{0.32\textwidth}
        \centering
        \includegraphics[width=\linewidth]{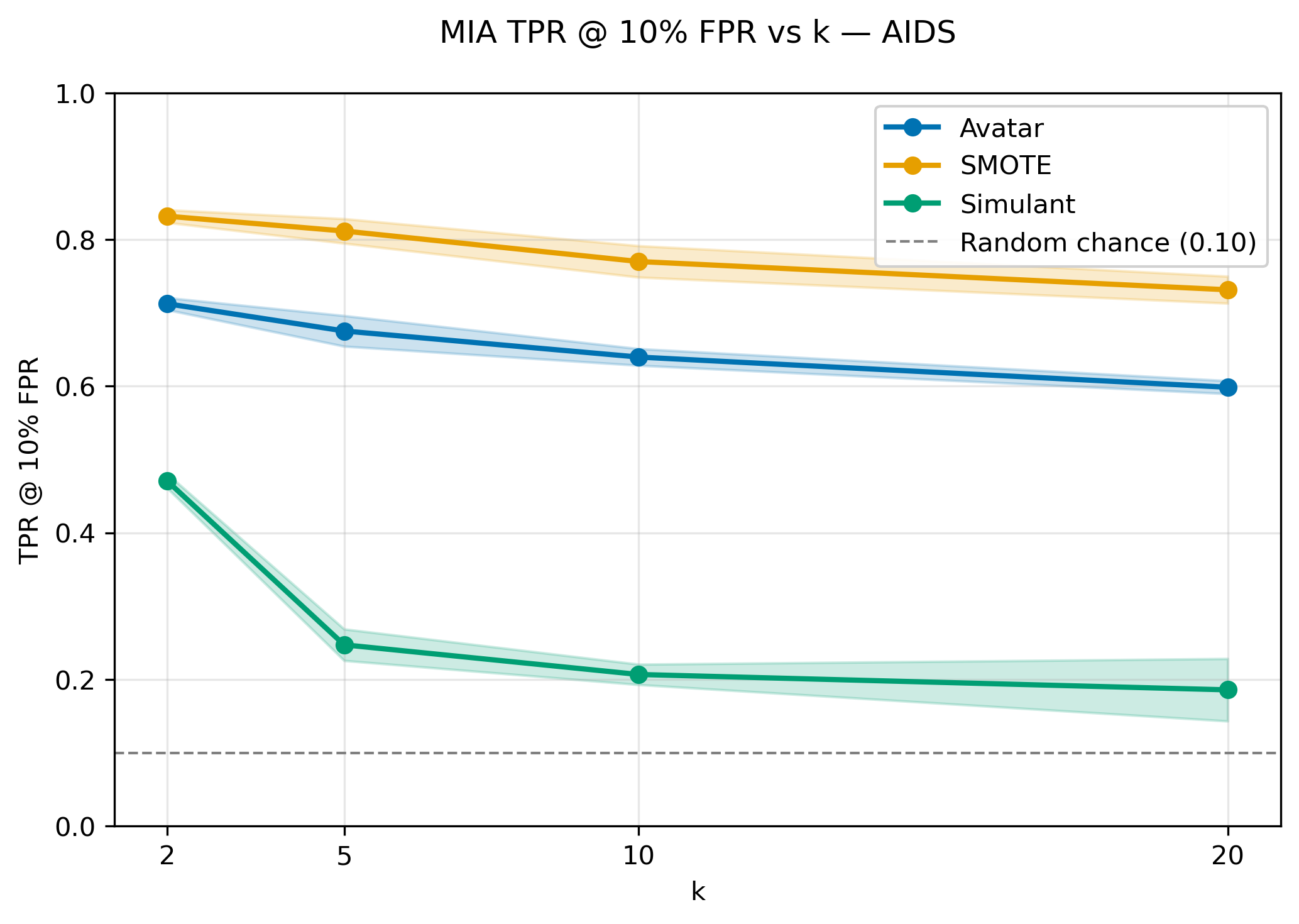}
        \caption{AIDS}
        \label{fig:linkage_aids}
    \end{subfigure}
    \hfill
    \begin{subfigure}{0.32\textwidth}
        \centering
        \includegraphics[width=\linewidth]{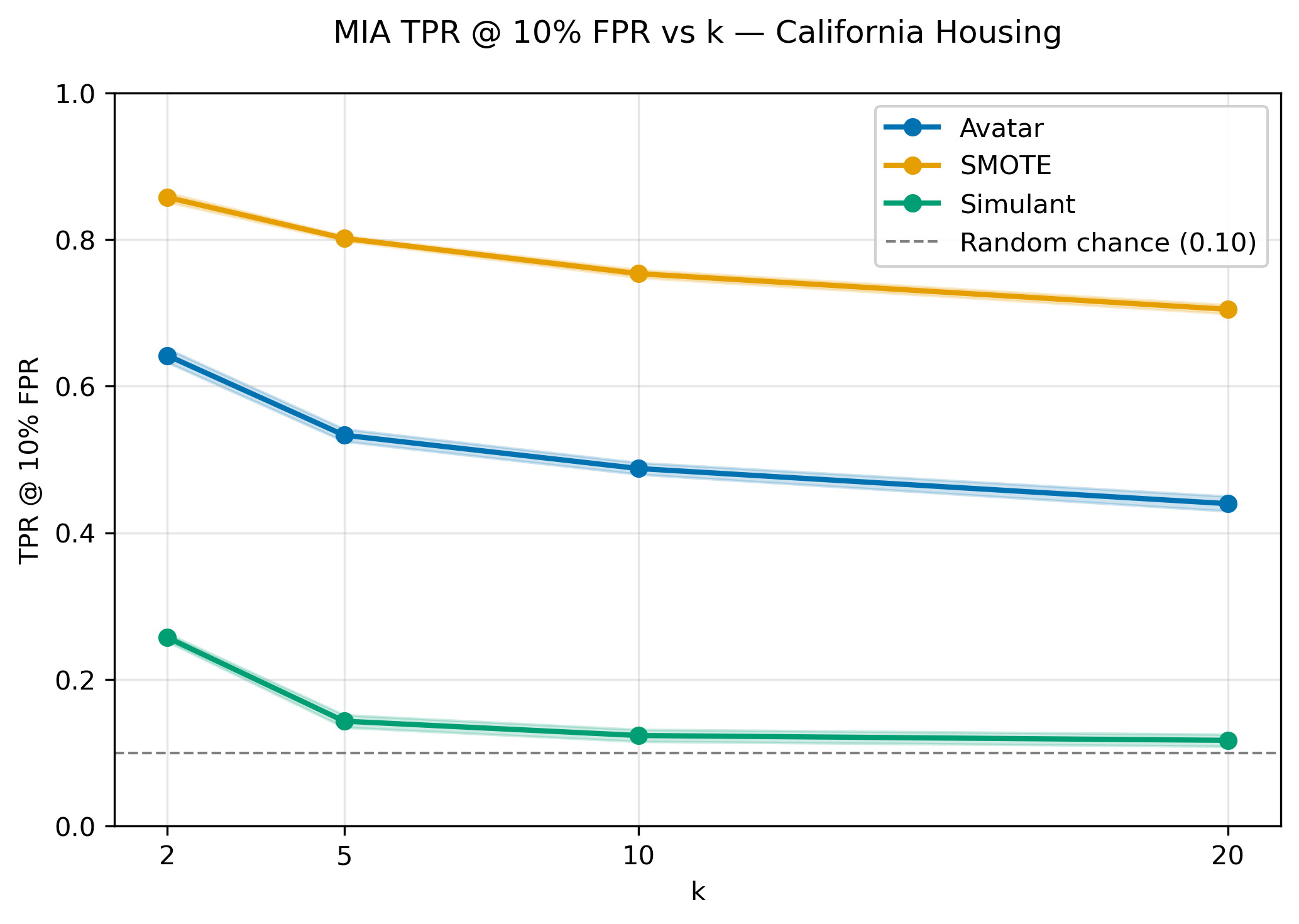}
        \caption{California Housing}
        \label{fig:linkage_california}
    \end{subfigure}

    \caption{MIA's TPR\@10\%FPR for against each \SLC~generative method for each dataset with 95\% CI}
    \label{fig:MIA_k}
\end{figure*}

Finally, Figure~\ref{fig:MIA_k} describes the MIA success in term of TPR@10\%FPR for varying values of $k$. 
As expected increasing the value of $k$ generally decreases the attack success rate. 
However, the privacy leakage measured by the MIA remains non-trivial for synthetic data generated by both SMOTE and Avatar with attack success rate four to eight time larger than the random guessing baseline.

\subsection{Likelihood-based linkage attack}

When evaluating linkage attacks between real and synthetic data points, a random-guess strategy would succeed with probability approximately $1/450$, $1/2000$ and $1/20000$, respectively for WBCD, AIDS and California Housing. 
As shown in Table~\ref{tab:linkage_k10}, the observed success rates are orders of magnitude higher than random guessing for all \SLC, with such substantial residual privacy linkage risk, one can doubt that those synthetic data can be consider anonymous.

\begin{table}[h!]
\centering
\caption{Linkage attack performance for $k=10$. Values are percentages (mean $\pm$ 95\% CI), ``Closest'' links each synthetic profile to the closest real profile in the original space while ``Lebrun et al.'' denotes the latent-space baseline. 
For each row, the best full-dataset result among \{Attack, Closest, Lebrun et al.\} and the best filtered result among \{Attack (top 10\%), Closest (10\%), Lebrun et al. (10\%)\} are highlighted in bold.}
\label{tab:linkage_k10}
\scriptsize
\setlength{\tabcolsep}{2.8pt}
\resizebox{\linewidth}{!}{%
\begin{tabular}{llcccccc}
\toprule
Dataset & Method & Attack & \makecell[c]{Attack\\(top 10\%)} & Closest & \makecell[c]{Closest\\(10\%)} & \makecell[c]{Lebrun\\et al.} & \makecell[c]{Lebrun et al.\\(10\%)} \\
\midrule
\multirow{3}{*}{WBCD}
 & Avatar   & \textbf{10.20 $\pm$ 1.05} & \textbf{8.79 $\pm$ 3.00}  & 0.45 $\pm$ 0.31 & 0.17 $\pm$ 0.13 & 0.98 $\pm$ 0.30 & 0.18 $\pm$ 0.14 \\
 & Simulant & \textbf{5.32 $\pm$ 0.59}  & \textbf{6.14 $\pm$ 1.79}  & 4.81 $\pm$ 0.37 & 5.63 $\pm$ 1.49 & 4.37 $\pm$ 0.47 & 5.84 $\pm$ 1.54 \\
 & SMOTE    & \textbf{47.53 $\pm$ 2.34} & \textbf{65.33 $\pm$ 7.37} & 43.79 $\pm$ 2.04 & 40.70 $\pm$ 2.99 & 26.59 $\pm$ 1.44 & 42.89 $\pm$ 2.90 \\
\midrule
\multirow{3}{*}{AIDS}
 & Avatar   & \textbf{29.23 $\pm$ 0.48} & \textbf{54.49 $\pm$ 2.63} & 4.24 $\pm$ 0.59 & 1.78 $\pm$ 0.79 & 4.93 $\pm$ 0.52 & 1.59 $\pm$ 1.18 \\
 & Simulant & \textbf{2.79 $\pm$ 0.13}  & 2.90 $\pm$ 1.17          & 1.84 $\pm$ 0.28 & 3.08 $\pm$ 0.85 & 1.80 $\pm$ 0.22 & \textbf{3.18 $\pm$ 0.98} \\
 & SMOTE    & \textbf{53.83 $\pm$ 1.55} & \textbf{85.61 $\pm$ 1.62} & 50.38 $\pm$ 1.23 & 52.99 $\pm$ 3.07 & 44.28 $\pm$ 1.10 & 53.36 $\pm$ 3.42 \\
\midrule
\multirow{3}{*}{\makecell[c]{California\\Housing}}
 & Avatar   & \textbf{21.47 $\pm$ 0.69} & \textbf{35.46 $\pm$ 1.11} & 6.41 $\pm$ 0.27 & 3.49 $\pm$ 0.52 & 17.95 $\pm$ 0.82 & 14.44 $\pm$ 1.01 \\
 & Simulant & \textbf{1.82 $\pm$ 0.11}  & \textbf{3.13 $\pm$ 0.45}  & 0.81 $\pm$ 0.06 & 1.23 $\pm$ 0.07 & 0.54 $\pm$ 0.04 & 0.31 $\pm$ 0.07 \\
 & SMOTE    & 35.35 $\pm$ 4.36          & 50.21 $\pm$ 9.69          & \textbf{49.76 $\pm$ 0.28} & \textbf{51.17 $\pm$ 0.55} & 35.40 $\pm$ 0.62 & 43.70 $\pm$ 1.44 \\
\bottomrule
\end{tabular}%
}
\end{table}

For synthetic profiles generated by Avatar, our likelihood-based linkage attack clearly outperforms both baselines on all three datasets. 
For $k=10$, the attack correctly links an avatar to its seed profile in $10$ to $30\%$ of the cases while both distance-based baselines remain substantially lower. 
This confirms that, for Avatar, modeling the seed-conditional distribution of synthetic points is more effective than relying solely on nearest-neighbor distances, whether in the original or latent space. 
The lower attack success rates with the WBCD dataset can be explained by the fact that it contains only categorical features, with the likelihood/density estimation working poorly for categorical features. 

Similar to the MIA results, SMOTE is the most vulnerable method. 
For $k=10$, our proposed attack successfully links synthetic profiles to their original seeds in one third to half of the cases depending on the dataset.
Baselines are already very strong and even better than our proposed approach for California Housing, which indicates that much of the leakage can already be captured by simple distance-based baseline.

For Simulant, our likelihood-based attack also achieves the best attack success rate on all three datasets, although the margin over the baselines is smaller. 
These values remain significantly above random guessing, but they also suggest that Simulant is less vulnerable than Avatar and SMOTE.
Moreover, measuring the percentage of correct associations found by the attack, both for all the data points and for the $10\%$ for which the likelihood ratio is the highest, we can observe that the attack success rate increases when attacking only profiles with the highest confidence. 
This observation suggests that likelihood-based approach is better fitted for identifying confident linkage.

\begin{figure*}[h!]
    \centering
    \begin{subfigure}{0.32\textwidth}
        \centering
        \includegraphics[width=\linewidth]{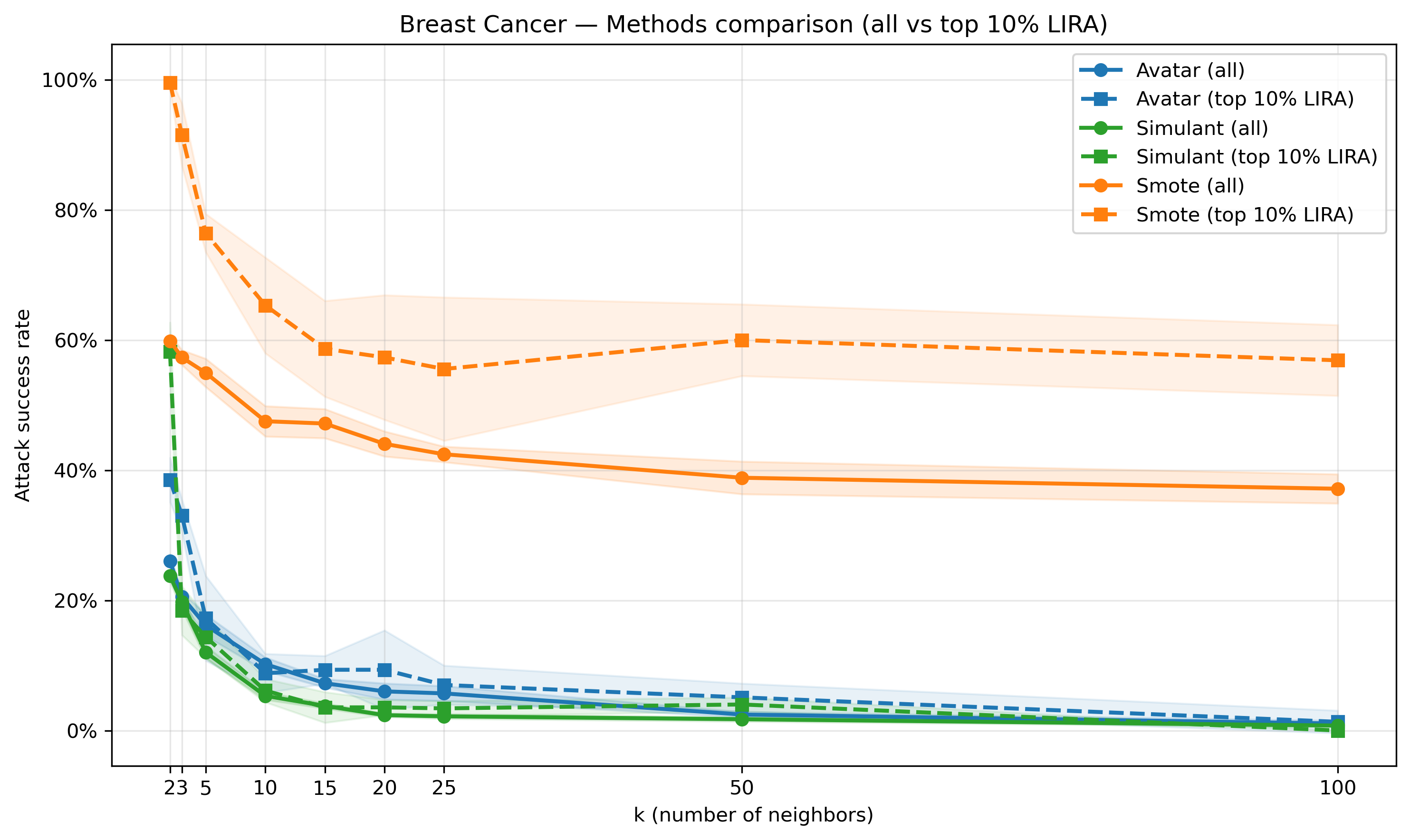}
        \caption{WBCD}
        \label{fig:linkage_breast}
    \end{subfigure}
    \hfill
    \begin{subfigure}{0.32\textwidth}
        \centering
        \includegraphics[width=\linewidth]{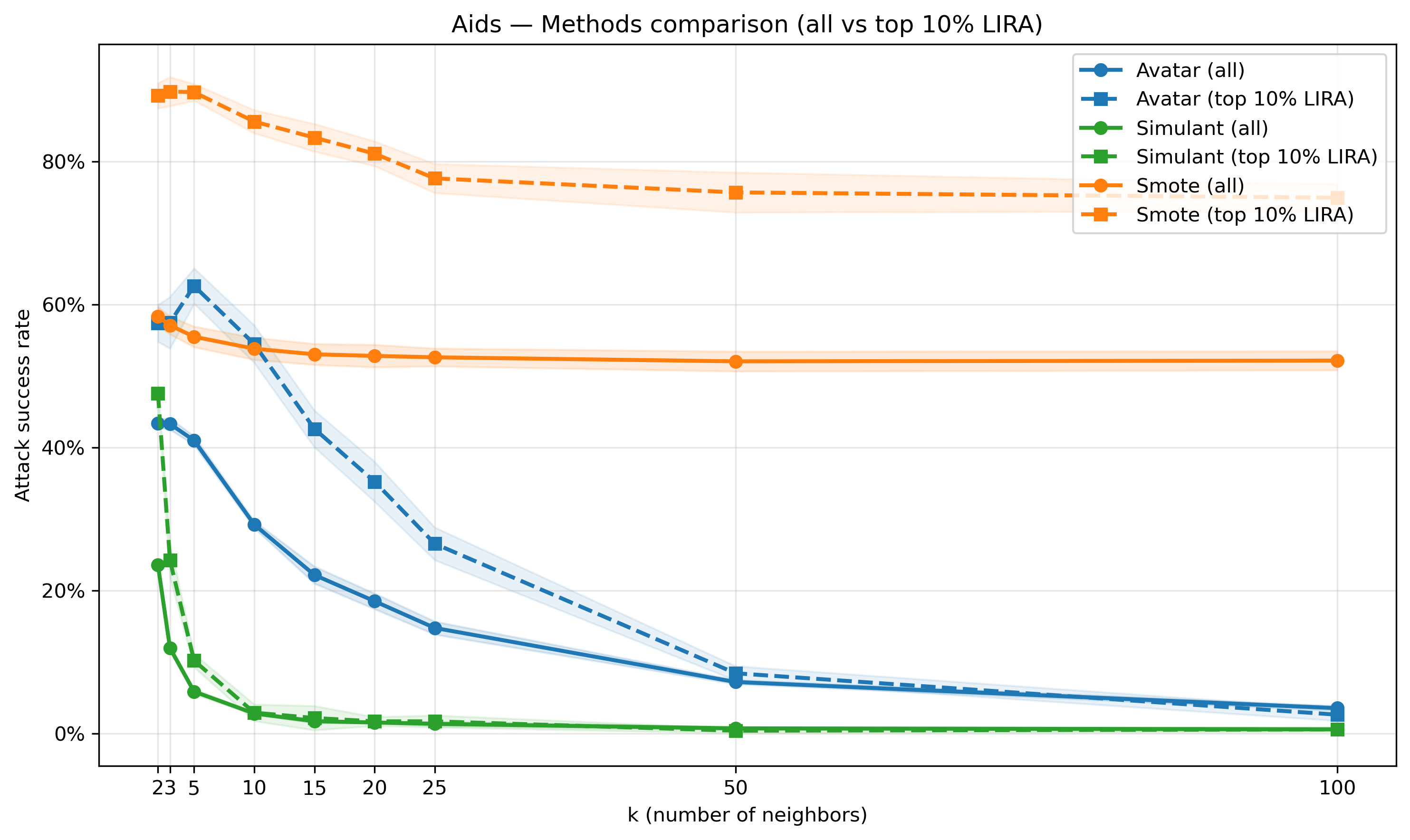}
        \caption{AIDS}
        \label{fig:linkage_aids}
    \end{subfigure}
    \hfill
    \begin{subfigure}{0.32\textwidth}
        \centering
        \includegraphics[width=\linewidth]{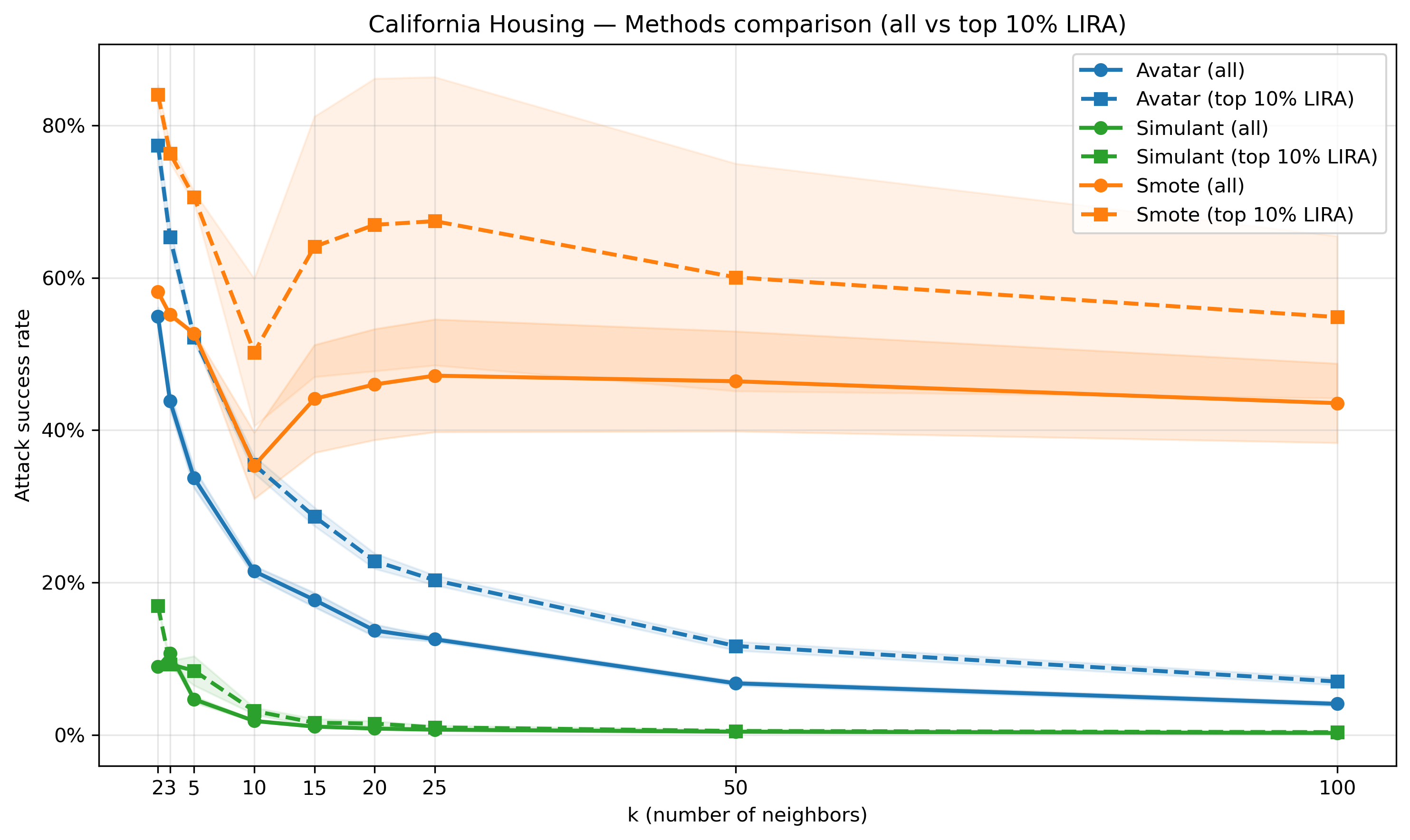}
        \caption{California Housing}
        \label{fig:linkage_california}
    \end{subfigure}

    \caption{Linkage attack against each \SLC~generative method for each dataset with 95\% CI. The plain line presents the overall linkage accuracy while the dotted line present accuracy for links with (top-10\%) highest likelihood ratio}
    \label{fig:linkage}
\end{figure*}

Figure \ref{fig:linkage} presents our likelihood-based linkage attack's success rate for different values of $k$.
As expected, increasing the value of $k$ generally decreases the attack success rate but nonetheless the synthetic data produced by SMOTE remain highly vulnerable to linkage attack.
In contrast, increasing the value of $k$ decrease the linkage success rate for synthetic data produced by Avatar. 
However, even for $k=100$ some observations remain vulnerable to linkage with an attack success rate of respectively 1.07, 3.53 and 4.06 \% for WBCD, AIDS and California Housing. 
Those values are much higher than random guessing and suggest that augmenting the value of $k$ may not be sufficient to protect real data from being linked to their synthetic counterparts.

\subsection{Reconstruction attack}
To assess the effectiveness of our proposed reconstruction attack against synthetic data produced by Avatar with $k=2$, we report the number of correctly reconstructed profiles as well as the precision, which is the percentage of reconstructed points corresponding to a real point. 
When evaluating if a reconstruction is correct, a small margin of error is tolerated as measured by the $L_1$ distance. 
This margin is defined as $\tau d$ in which $d$ is the dimensionality and $\tau$ is a dataset specific threshold, which can be interpreted as the average absolute error allowed by column. 
Since the range and nature of variables differ from one dataset to another, we adapt the parameter to the characteristics of the dataset. 
More precisely, the value of $\tau$ is set respectively to 0.01, 0.5 and 0.1 respectively for WBCD, AIDS and California Housing. 
The chosen tolerance is verified to be reasonable by validating that (almost) no two real profiles are closer than $d \tau$ (0.02\% for California Housing). 

Table~\ref{tab:reconstruction_freq_filter} presents the results of our reconstruction attack, which reconstructs one quarter of the real dataset from the synthetic data points while achieving a 80\% precision after filtering on the WBCD dataset. 
The baseline \textsf{ReconSMOTE} attack cannot reconstruct any point for both AIDS and California Housing dataset while it reconstructs 19 points with 95\% precision for WBCD. 
While our approach demonstrates more limited success on AIDS and California Housing, it still reconstructs a non-negligible number of real observations and our frequency filter allows us to boost the precision. 
Table 7
in Appendix E 
presents the performance of \textsf{ReconSMOTE} and our proposed reconstruction attack for synthetic data produced by SMOTE with no data augmentation.

\begin{table}[h!]
\centering
\small
\setlength{\tabcolsep}{5pt}
\renewcommand{\arraystretch}{1.05}
\caption{Reconstruction precision with and without frequency filtering.}
\label{tab:reconstruction_freq_filter}
\begin{tabular}{llcc}
\toprule
\textbf{\multirow{2}{*}{Dataset}} &  & \textbf{\multirow{2}{*}{\makecell{Number of correct \\ reconstructed points}}} & \textbf{\multirow{2}{*}{Precision}} \\
\\
\midrule
\multirow{2}{*}{WBCD}
& all reconstructions & 141 & 9\% \\
& freq filter = 5 & 122 & 80\% \\
\midrule
\multirow{2}{*}{AIDS}
& all reconstructions & 17 & 3\% \\
& freq filter = 5 & 10 & 4.5\% \\
\midrule
\multirow{2}{*}{\makecell{California\\Housing}}
& all reconstructions & 129 & 2.6\% \\
& freq filter = 5 & 17 & 6\% \\
\bottomrule
\end{tabular}%
\end{table}

\section{Differential privacy and \SLC}
\label{sec:discussion}

Our work shows that all three SLC methods studied generate synthetic data that are susceptible to various kinds of privacy attacks. 
A potential solution would be to modify these methods to ensure that they satisfy differential privacy (DP).
In particular, DP bounds the maximum contribution a single observation can have to the output of a randomized algorithm. 
As stated before, the authors of the Simulant method claim that their proposed \SLC~generative method achieves $\varepsilon$-DP. 
We do not agree with their claim as

Simulant assumes direct access to the data to choose a seed and to compute the distance to all other real profiles for neighbor selection, with the noise only added in the combination step. 
For an \SLC~ algorithm to be considered differentially-private, all steps of pipeline must be taken into account in the privacy analysis, with relevant noise being added at each step.
Moreover, the noise addition methodology itself is not appropriate to satisfy DP. 
Indeed, the Gaussian mechanism used to achieve DP relies on noise addition while Simulant uses multiplicative Gaussian noise. 
In general, this does not achieved DP, in particular when the protected quantity can take the value zero. 
In addition, DP requires the scale of the noise distribution to be calibrated to the sensitivity of the released quantity, whereas Simulant uses a fixed quantity.

A $\varepsilon$-DP version of SMOTE has also been proposed \cite{lut2022privacy}. 
This approach relies on projecting real data onto a multi-dimensional hyper-grid, counting the number of observation falling in each bin and adding noise to it. 
Grid cell centers then act as proxy for observations when performing the SMOTE algorithm.
More precisely instead of using an observation as a seed, the algorithm randomly selects a cell center to act as seed for generation, with probability depending on the number of observations in this cell, with noise carefully added in this step to ensure DP. 
Thus, while the DP property of this method is correct, the proposed methodology can no longer be considered part of the \SLC~family of method since no real profile acts as a seed, neither as a neighbor. 
We leave as future work the possibility of designing a truly differentially-private \SLC, although we posit that such mechanism may not provide much utility due to the addition of noise to each step accounting for the worst case in which a given observation is repeatedly selected as a neighbor.

\section{Conclusion}
\label{sec:conclusion}

While being less notorious than deep-learning base generative techniques, \SLC~generation approaches have been used in real-world applications with high stakes such as healthcare. 
Our work provides extensive empirical privacy evaluation of three such methods, demonstrating in particular significant residual privacy risks through different types of attacks. 
In particular, our attacks include an enhanced MIA attack based on ensemble learning, a novel linkage attack designed for \SLC~methods as well as a reconstruction attack against synthetic data generated by Avatar with privacy parameter $k=2$.
These attacks have success rates often an order of magnitude higher than random guessing, indicating clear privacy leakage. 
Additionally, our experiments illustrate once again that post-hoc distance-based metrics are poor representative of the real privacy risk, thus leading to a false sense of privacy, as shown also in recent works~\cite{ganev2025inadequacy, yao2025dcr}. 
In particular, our likelihood-based linkage attack still succeeds in situation in which post-hoc metrics indicate a low risk. 
For example, our proposed linkage attack achieves 18\% success rate on synthetic data produced by the Avatar method with $k=20$ while the privacy scores did not suggest any significant privacy leakage.
Even worse on WBCD for synthetic data generated with Avatar $k=2$ (a setting used for real open data release), post-hoc metrics do not suggest any significant risk but our reconstruction attack can reconstruct with high precision one quarter of the original private observations from synthetic data points only.

\section*{Responsible disclosure approach}

We acknowledge that our research focuses on privacy attacks against privacy techniques designed to be used for real and sensitive use-cases. 
However, we believe that highlighting those privacy risks will demonstrate that they need to be used with caution. 
In contrast, not publishing our findings could lead to increased privacy risks as people may have a false sense of privacy while malicious actors could still find and exploit the same vulnerabilities. 
In particular, synthetic profiles from health data from real person have already been published in open data.
As it is impossible to ensure that all potentially vulnerable copy of this data has been deleted, we will not open source the code relative to the reconstruction attack. 
We have also been engaged in a responsible disclosure process, sharing in particular our findings with companies whose products are the subject to this research and waiting a standard 90 days before publishing the results.
One of the contacted companies has shown interest in the our approach and we had several meetings with it to provide more details about our attacks.

\bibliographystyle{splncs04}
\bibliography{references}

\appendix

\section{Attribute Inference Attack}
\label{Appendix:AIA}
\subsection{Attribute Inference Attack Description}

Attribute inference attacks (AIAs) are a type of privacy attacks in which an adversary has access to a part of the profile of a target individual known to participate to the training dataset (\emph{e.g.}, a subset of attribute values). 
The objective of the adversart is to infer the unknown part of the profile based on the synthetic dataset. 
For instance, the adversary could have access to all attributes except one and tries to predict it. 
Our attack works in the no-box setting (no adversarial access to the model) and without additional background knowledge on the original dataset.
However, one must be careful when measuring the privacy leakage through attribute inference attacks in order to ensure that it does not just learn how to perform an imputation on a missing attribute from the available data~\cite{jayaraman2022attribute}.
As a result, we compare the attack success rate to a baseline learnt on an auxiliary dataset. 
More precisely, 
similarly to related works~\cite{giomi2022unified, octopize_aia}, we rely on a $k$-nearest neighbour classifier ($k$-NN) to perform the attribute inference attack.

Algorithm~\ref{alg:aia_knn} describes the attack pipeline. 
The attack works by looping over all attributes, considering them one by one as the missing attribute the adversary tries to infer. 

The original dataset is split in half between the guessing set and the auxiliary dataset. 
Then, two $k$-NN classifiers with $k=5$ are trained, one on synthetic data points obtained from the guessing set and another on the auxiliary data. 
Afterwards, the percentage of attributes for which the adversary has a better inference performance using the synthetic dataset rather than the auxiliary dataset is computed. 
Relying on a $k$-NN for the attribute inference attack allows to measure the proportion of features in the dataset for which the synthetic values of the features are closer to original data points than normally expected. 
In such cases, the synthetic data reveal more information about the missing attributes than the mere knowledge of the data distribution, thus highlighting a privacy issue. 
More precisely, quantifying the privacy leakage with this attack requires access to an auxiliary set of observation from the distribution. 
However, this auxiliary dataset is not used by the adversary but rather serve as a way to ensure that the attack measures privacy leakage beyond simple imputation that can be learnt from data itself.

\begin{algorithm}[h!]
  \caption{Attribute Inference Attack}
  \label{alg:aia_knn}

  \SetKwFunction{KNNTrain}{KNN\_Train}
  \SetKwFunction{Predict}{Predict}

  \Input{
    Real dataset $D=\{x_1,\dots,x_n\}$ with $d-1$ attributes,
    synthetic dataset $S$  generated from $D$ (with all $d$ features), feature $j$ to be inferred by attacker}
  \Output{
    Predicted feature value for the feature unknown to the attacker
  }

\BlankLine
\textbf{Train attack model on synthetic data to predict feature $j$.}\\
$C^{S}_j \leftarrow \KNNTrain(S) $\;

    \BlankLine
    \textbf{Predict missing feature value}\\
      $\hat{x}^{S}_{j}(x) \leftarrow \Predict(D)$\;

  \Return $\hat{x}^{S}_{j}(x)$ ;
\end{algorithm}

\subsection{Attribute Inference Attack Results}

Table~\ref{tab:inference_power} shows for each dataset and every \SLC~generative method the percentage of columns in which  synthetic data provides a better inference power than auxiliary data from the data distribution.
A percentage higher than $50\%$ indicates that the use of synthetic data leads to a better attribute inference accuracy than auxiliary data coming from the same distribution, which indicates a privacy leakage. 
SMOTE is extremely vulnerable to this attack. 
For almost all features and across all datasets, synthetic data yields more inference power against the original data compared to auxiliary data.
With slightly less inference power, synthetic data produced by Avatar are also vulnerable to attribute inference attack.
Finally, Simulant seems to be much more resilient against this attack with a majority of features that can be better predicted with auxiliary data than synthetic data.

\begin{table}[h!]
\centering
\caption{Attribute inference attack: global performance (\%) by dataset and \SLC~generative method with $k=10$.}
\label{tab:inference_power}
\begin{tabular}{lccc}
\hline
\textbf{Dataset} & \textbf{Avatar} & \textbf{SMOTE} & \textbf{Simulant} \\
\hline
WBCD     & 100.0 & 100.0 & 70.0 \\
AIDS               & 88.0  & 100.0 & 28.0 \\
California\_housing & 77.78 & 88.89 & 0.0 \\
\hline
\end{tabular}
\end{table}

Recall that the attribute inference attack is conducted using a $k$-NN. 
Results can be interpreted for both SMOTE and Avatar as synthetic data being closer than auxiliary data to the true value of real original data. 
This is consistent with the process producing the synthetic profiles for these methods, which interpolates between the real values of neighbors.

\section{Data restricted linkage attack}
\subsection{Data restricted linkage attack setting}
\label{Appendix:Linkag_restricted}

The generic linkage attack presented in the main paper body can be considered unrealistic since it assumes that the adversary has access to both the real dataset and the synthetic one.
A more realistic setting would be one in which the adversary only has access to one real profile $x_{seed}$ and the set of synthetic observations $S$.
In this setting, the adversary chooses uniformly at random a real profile $x$ from $D$. 
Then, given access to the synthetic profiles, the attacker tries to identify the synthetic profile $s_{seed}$ that originates from $x$ being selected as the seed. 

Algorithm~\ref{alg:linkage_restricted} describes the corresponding linkage, which differs from the previous one in several ways. 
First, the adversary uses the synthetic dataset instead of the original one to simulate synthetic observations obtained from the chosen point $x$. 
If the synthetic observations are distributed similarly to the original ones, the distributions obtained in this manner will be similar to those than a more knowledgeable adversary could obtain using the entire original dataset.  
Afterwards, instead of looking at the probability of a target profile $s$ coming from the distribution obtained from profile $x$, the adversary computes the log likelihood of each synthetic profile coming from the distribution of simulated synthetic data points obtained from $x$. 
Similarly, the likelihood ratio is computed between the first and second synthetic profile candidates as a score for the linkage.
Note that the same baselines are used here as in the generic linkage attack.

\begin{algorithm}[h!]
  \caption{Data-restricted linkage attack via likelihood modeling}
  \label{alg:linkage_restricted}

  \SetKwFunction{Gen}{SLC\_Generate}
  \SetKwFunction{ModelDist}{ModelDistribution}
  \SetKwFunction{Lik}{LogLikelihood}
  \SetKwFunction{KNN}{kNN}

  \Input{
    One real observation $x_{\mathrm{seed}}$, released synthetic dataset $S$,
    \SLC generator $\Gen(\cdot)$, number of simulations $t$,
    candidate set size $k'$, distance function $\mathrm{dist}(\cdot,\cdot)$.
  }
  \Output{
    Link score $\mathrm{score}(x_{\mathrm{seed}})$ and predicted link $\widehat{s}(x_{\mathrm{seed}})\in S$.
  }

  \BlankLine
  $\ModelDist(\{s^{(1)},\dots,s^{(t)}\})$ fits a parametric or non-parametric model for the seed-conditional
  distribution of synthetic outputs, which for Avatar corresponds to estimate $(\mu_{\mathrm{seed}}, \Sigma_{\mathrm{seed}})$ while for SMOTE/Simulant it refers to fitting a Gaussian KDE.

  \BlankLine
  $\Lik(s \mid \mathcal{M}_{\mathrm{seed}})$ returns the log-likelihood (or density) of $s$ under the fitted model
  $\mathcal{M}_{\mathrm{seed}}$.

  \BlankLine
  \textbf{Step 1: build the seed-conditional synthetic-output distribution using $S$.}\\
  \For{$r \leftarrow 1$ \KwTo $t$}{
    $S^{(r)} \leftarrow \Gen(S)$\;
    Extract $s^{(r)}_{\mathrm{seed}}$ as the synthetic output generated from seed $x_{\mathrm{seed}}$ in simulation $r$\;
  }
  $\mathcal{S}_{\mathrm{seed}} \leftarrow \{\, s^{(r)}_{\mathrm{seed}} : r \in \{1,\dots,t\}\,\}$\;
  $\mathcal{M}_{\mathrm{seed}} \leftarrow \ModelDist(\mathcal{S}_{\mathrm{seed}})$\;

  \BlankLine
  \textbf{Step 2: identify the linked synthetic record in $S$.}\\
  $C(x_{\mathrm{seed}}) \leftarrow \KNN(S, x_{\mathrm{seed}}, k')$\;

  \ForEach{$s \in C(x_{\mathrm{seed}})$}{
    $\ell(s \mid x_{\mathrm{seed}}) \leftarrow \Lik(s \mid \mathcal{M}_{\mathrm{seed}})$\;
  }

  Let $s^\star(x_{\mathrm{seed}}) \leftarrow \arg\max_{s \in C(x_{\mathrm{seed}})} \ell(s \mid x_{\mathrm{seed}})$\;
  Let $s^{(2)}(x_{\mathrm{seed}})$ be the second-best candidate in $C(x_{\mathrm{seed}})$ according to $\ell(s \mid x_{\mathrm{seed}})$\;

  $\widehat{s}(x_{\mathrm{seed}}) \leftarrow s^\star(x_{\mathrm{seed}})$\;
  $\mathrm{score}(x_{\mathrm{seed}}) \leftarrow
    \log \ell(s^\star(x_{\mathrm{seed}}) \mid x_{\mathrm{seed}})
    - \log \ell(s^{(2)}(x_{\mathrm{seed}}) \mid x_{\mathrm{seed}})$\;

  \Return{$\widehat{s}(x_{\mathrm{seed}}), \mathrm{score}(x_{\mathrm{seed}})$}\;
\end{algorithm}

\subsection{Data restricted linkage attack results}

\begin{table}[h!]
\centering
\caption{Average data-restricted linkage attack performance for $k=10$ over 100 randomly selected targets. Values are percentages (mean $\pm$ 95\% CI), ``Closest'' denotes the original-space baseline and corresponds to ``Baseline'', while ``Lebrun et al.'' denotes the latent-space baseline and corresponds to ``Baseline latent''. For each row, the best full-dataset result among \{Attack, Closest, Lebrun et al.\} and the best filtered result among \{Attack (top 10\%), Closest (10\%), Lebrun et al. (10\%)\} are highlighted in bold.}
\label{tab:privacy_game_k10_summary_random_only}
\scriptsize
\setlength{\tabcolsep}{2.8pt}
\resizebox{\linewidth}{!}{%
\begin{tabular}{llcccccc}
\toprule
Dataset & Method & Attack & \makecell[c]{Attack\\(top 10\%)} & Closest & \makecell[c]{Closest\\(10\%)} & \makecell[c]{Lebrun\\et al.} & \makecell[c]{Lebrun et al.\\(10\%)} \\
\midrule
\multirow{3}{*}{WBCD}
 & Avatar   & \textbf{2.60 $\pm$ 1.40}  & \textbf{2.00 $\pm$ 1.24}  & 1.00 $\pm$ 0.87  & 0.00 $\pm$ 0.00  & \textbf{2.60 $\pm$ 1.40}  & 0.00 $\pm$ 0.00 \\
 & Simulant & \textbf{2.80 $\pm$ 1.45}  & \textbf{4.00 $\pm$ 1.74}  & 2.40 $\pm$ 1.34  & 2.00 $\pm$ 1.24  & 2.40 $\pm$ 1.34  & 2.00 $\pm$ 1.24 \\
 & SMOTE    & \textbf{48.40 $\pm$ 4.38} & \textbf{84.00 $\pm$ 3.25} & 34.40 $\pm$ 4.17 & 52.00 $\pm$ 4.42 & 46.20 $\pm$ 4.37 & 46.00 $\pm$ 4.41 \\
\midrule
\multirow{3}{*}{AIDS}
 & Avatar   & 17.00 $\pm$ 3.30 & \textbf{32.00 $\pm$ 4.13} & 2.00 $\pm$ 1.23  & 0.00 $\pm$ 0.00  & \textbf{19.20 $\pm$ 3.46} & 2.00 $\pm$ 1.24 \\
 & Simulant & 1.20 $\pm$ 0.96  & 2.00 $\pm$ 1.24  & \textbf{3.20 $\pm$ 1.54}  & \textbf{14.00 $\pm$ 3.07} & 2.20 $\pm$ 1.29  & 2.00 $\pm$ 1.24 \\
 & SMOTE    & \textbf{58.00 $\pm$ 4.33} & \textbf{84.00 $\pm$ 3.25} & 29.80 $\pm$ 4.01 & 56.00 $\pm$ 4.40 & 57.80 $\pm$ 4.33 & 58.00 $\pm$ 4.37 \\
\midrule
\multirow{3}{*}{\makecell[c]{California\\Housing}}
 & Avatar   & \textbf{24.40 $\pm$ 3.77} & \textbf{38.00 $\pm$ 4.30} & 0.40 $\pm$ 0.55 & 0.00 $\pm$ 0.00  & 18.80 $\pm$ 3.43 & 4.00 $\pm$ 1.74 \\
 & Simulant & 0.20 $\pm$ 0.39  & {0.00 $\pm$ 0.00}  & 0.20 $\pm$ 0.39 & {0.00 $\pm$ 0.00}  & \textbf{1.00 $\pm$ 0.87}  & {0.00 $\pm$ 0.00} \\
 & SMOTE    & 26.20 $\pm$ 3.86 & \textbf{70.00 $\pm$ 4.06} & 5.20 $\pm$ 1.95 & 20.00 $\pm$ 3.54 & \textbf{54.00 $\pm$ 4.37} & 38.00 $\pm$ 4.30 \\
\bottomrule
\end{tabular}%
}
\end{table}

The experiment was conducted independently on 100 randomly selected points by repeating the experiment (\emph{i.e.}, the generation of a fresh batch of synthetic data points) five times. 
Similar to the generic linkage attack, we measure how higher link prediction score translate to the attack success.
Table~\ref{tab:privacy_game_k10_summary_random_only} presents the attack and baseline success rates for each dataset for a fixed value of $k=10$. 
Both the likelihood-based attack and Lebrun et al. baseline produced similar results when evaluating all guesses. However, when the attacker focuses on the 10\% of links predicted with highest confidence our proposed linkage-based attack outperforms both baselines.  
With the exception of synthetic data produced by Simulant on California Housing, the obtained results are significantly (\emph{i.e.}, often at least an order of magnitude) better than random guessing with particular high performance against SMOTE and AVATAR evaluated on AIDS and California Housing. 

The likelihood attack performance dropped in this data restricted set-up compared to the generic one. 
This performance degradation can be attributed to the fact that estimating distribution of synthetic data points using synthetic data only leads to a poorer estimation of the distribution. 

\subsection{Reconstruction Attack against SMOTE}
\label{Appendix:reconstruction}

\begin{table}[h!]
\centering
\small
\caption{Reconstruction attack results against synthetic data produced by SMOTE with and without frequency filtering.}
\label{tab:reconstruction_smote}
\begin{tabular}{llcc}
\toprule
\textbf{Dataset} & \textbf{Setting} & \textbf{\makecell{Number of correct\\reconstructed points}} & \textbf{Precision} \\
\midrule
\multirow{2}{*}{WBCD}
& all reconstructions & 170 & 61\% \\
& freq.\ filter $=5$ & 115 & 72\% \\
\midrule
\multirow{2}{*}{AIDS}
& all reconstructions & 5 & 0.9\% \\
& freq.\ filter $=5$ & 5 & 2\% \\
\midrule
\multirow{2}{*}{\makecell{California\\Housing}}
& all reconstructions & 174 & 3\% \\
& freq.\ filter $=5$ & 49 & 3.5\% \\

\bottomrule
\end{tabular}
\end{table}

Table \ref{tab:reconstruction_smote} presents results of our proposed reconstruction attack against SMOTE with k=10. 
While the \textsf{ReconSMOTE} baseline can only reconstruct 66 observation with 75\% precision for the WBCD dataset, our approach reconstructs twice as much observation after filtering for WBCD and can still reconstruct some observations for AIDS ans California Housing datasets.

\subsection{Impact of number of kept feature in latent space for Avatar}
\label{Appendix:NCP}

\begin{figure}[h!]
        \centering
        \vspace{-3mm}
        \includegraphics[width=0.9\linewidth]{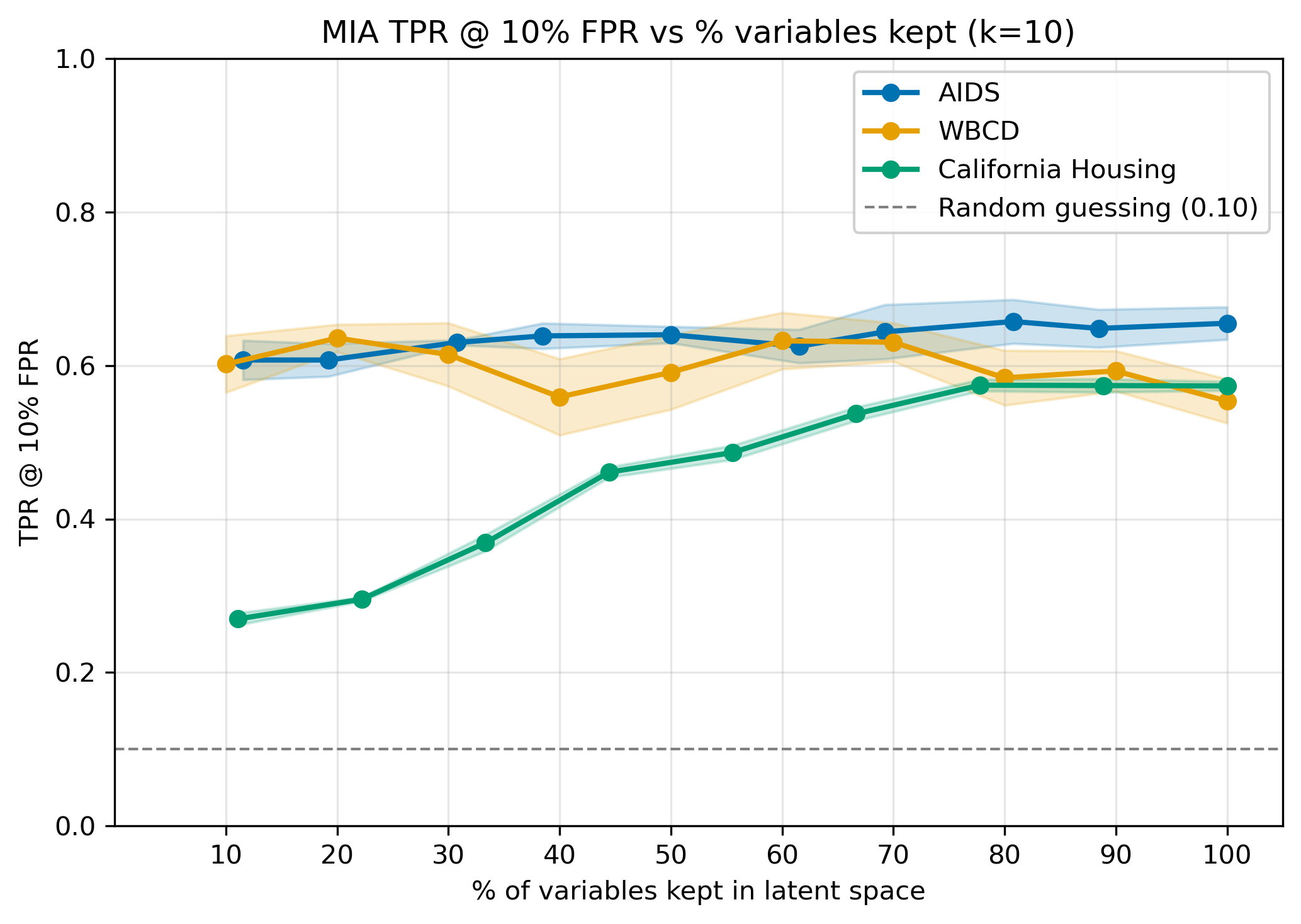}
        \caption{Illustration of the linkage attack using multivariate normal distributions.}
        \label{fig:ncp_effect}
\end{figure}

In Avatar, $k$ is the main privacy parameter while the number of features in the latent space used to identify the neighbors is being described as an additional hyper-parameter both in the original paper \cite{guillaudeux2023patient} as well as follow-up work \cite{demuth2025privacy}. 
Figure \ref{fig:ncp_effect} presents the impact of the number of kept principal components on the success (measured as TPR@10\%FPR) of our proposed ensemble MIA for a fix value of $k=10$.
For the MIA against synthetic data produced from the WBCD and AIDS dataset, the number of features in the latent space used to identify neighbors does not have an impact on the attack success. 
For the California Housing dataset, this parameter has a significant impact on the attack success. 
It is not clear why the number of features used to select the neighbors (while all features are used to compute the noisy weighted barycenter) has such an impact on the attack success rate. 
We leave this investigation as future work.

\end{document}